\documentclass[reprint,superscriptaddress,prd,nofootinbib]{revtex4-2}
\pdfoutput=1

\usepackage{graphicx}    
\usepackage{microtype}   
\usepackage{amsmath}     
  \usepackage{amsthm}    
  \usepackage{amssymb}   
 \usepackage{mathdots}
\usepackage{siunitx}     
\usepackage{booktabs}    
\usepackage[colorlinks,citecolor=blue]{hyperref}
\usepackage{array}

\usepackage{adjustbox}
\usepackage{mathtools}
\usepackage{slashed}
\usepackage[ruled,vlined]{algorithm2e}
\usepackage{tikz}
    \usetikzlibrary{math}
    \usetikzlibrary{decorations.pathreplacing}
    \usetikzlibrary{quantikz}

\usepackage{caption}
\usepackage{subcaption}

\newcommand{\bO}{\mathcal{O}}
\DeclarePairedDelimiter{\ceil}{\lceil}{\rceil}

\usepackage[capitalize]{cleveref}    
  \crefname{algorithm}{Alg.}{Algs.}
  \crefname{tab}{Table}{Tables}
  \crefname{fig}{Fig.}{Figs.}
  \crefname{section}{Sec.}{Secs.}
  \crefname{appendix}{App.}{Apps.}

\begin{document}

\title{Improving Quantum Simulation Efficiency of \\ Final State Radiation with Dynamic Quantum Circuits}

\author{Plato Deliyannis}
\email{pdeliyannis@lbl.gov}
\affiliation{Physics Division, Lawrence Berkeley National Laboratory, Berkeley, CA 94720, USA}

\author{James Sud}
\email{jamessud@berkeley.edu}
\affiliation{Department of Physics, UC Berkeley, Berkeley, CA 94720, USA}

\author{Diana Chamaki}
\email{dchamaki@berkeley.edu}
\affiliation{Department of Physics, UC Berkeley, Berkeley, CA 94720, USA}

\author{Zo\"{e} Webb-Mack}
\email{zwebbmack@lbl.gov}
\affiliation{Physics Division, Lawrence Berkeley National Laboratory, Berkeley, CA 94720, USA}

\author{Christian W. Bauer}
\email{cwbauer@lbl.gov}
\affiliation{Physics Division, Lawrence Berkeley National Laboratory, Berkeley, CA 94720, USA}

\author{Benjamin Nachman}
\email{bpnachman@lbl.gov}
\affiliation{Physics Division, Lawrence Berkeley National Laboratory, Berkeley, CA 94720, USA}

\date{\today}
\begin{abstract}
Reference~\cite{qps} recently introduced an algorithm (QPS) for simulating parton showers with intermediate flavor states using polynomial resources on a digital quantum computer.  We make use of a new quantum hardware capability called \textit{dynamical quantum computing} to improve the scaling of this algorithm, which significantly improves the method precision.  In particular, we modify the quantum parton shower circuit to incorporate mid-circuit qubit measurements, resets, and quantum operations conditioned on classical information.  This reduces the computational depth from $\bO(N^5\log_2(N)^2)$ to $\bO(N^3\log_2(N)^2)$ and the qubit requirements from $\bO(N\log_2(N))$ to $\bO(N)$. Using ``matrix product state'' statevector simulators, we demonstrate that the improved algorithm yields expected results for 2, 3, 4, and 5-steps of the algorithm. We compare absolute costs with the original QPS algorithm, and show that dynamical quantum computing can significantly reduce costs in the class of digital quantum algorithms representing quantum walks (which includes the QPS). Python code that implements QPS, both with and without dynamic gates, is publicly available on \href{https://github.com/LBNL-HEP-QIS/QPS_public}{Github}.
\end{abstract}

\maketitle

\section{Introduction}
\label{sec:intro}

High energy physics (HEP) simulations are one of the most natural and exciting applications of quantum computers, given the complex many-body quantum nature of HEP processes.  Foundational work establishing the existence of polynomial scaling digital quantum algorithms for scattering calculations~\cite{Jordan:2012xnu} has been followed by a variety of particle and nuclear physics investigations into simulations on quantum computers\footnote{There are now over a hundred papers in this area; see Ref.~\cite{Preskill:2018fag} for an overview to this topic and the papers that cite Ref.~\cite{Jordan:2012xnu} for specific studies.}.  

While most of these studies propose quantum algorithms for the full scattering process on a quantum computer, a complementary approach has been proposed to exploit factorization~\cite{Bauer:2021gup}.  In particular, scattering cross sections approximately factor into pieces governed by physical processes occurring at different energy scales.  One way to factorize a full calculation at a collider like the Large Hadron Collider (LHC) involves parton shower (PS) Monte Carlo.  Parton showers govern the collinear radiation from high energy charged particles.  Classical PSs approximate this radiation as a Markov Chain.  This is an excellent approximation in some cases, but ignores certain interference effects.  Recently, Ref.~\cite{qps} introduced a quantum algorithm for parton showers (QPS) that models interference effects from intermediate flavor states.  This algorithm requires only polynomial resources to model the same physics as existing exponentially scaling algorithms.  While the QPS does not describe the full properties of PSs in the Standard Model (dominated mostly by the strong force), it represents an important benchmark for developing and testing HEP algorithms on quantum computers.

Even though the QPS requires only polynomial quantum resources, it is still challenging to run on existing devices.  This is because we are in the Noisy Intermediate Scale Quantum (NISQ)~\cite{nisq} computing era where qubit counts, connectivities, and coherence times are limited, and quantum gate and readout operations have significant noise.  Therefore, there is a strong motivation to improve the polynomial scaling of the current quantum algorithms like QPS.  

In this paper, we improve the original QPS algorithm \cite{qps} by using \textit{dynamical quantum circuits}: we incorporate mid-circuit qubit measurements\footnote{Throughout the paper, we use the terms 'mid-circuit measurements' and 'remeasurement' interchangeably.} and quantum gates applied dynamically based on results of classical processing on these measurements.  The resulting quantum state prepared by the modified protocol is equivalent to the original. By re-setting the measured qubits to the ground state and re-using them for subsequent iterations, the computational complexity in both qubits and gates is reduced.  Compared to the original algorithm, the new version uses a factor of $\bO(N^2)$ fewer standard entangling gates, where $N$ is the number of points used to discretize the PS.

This paper is organized as follows.  Section~\ref{sec:dynamic} introduces dynamic quantum computing, in which a quantum processing unit (QPU) interacts with a classical processing unit (CPU) \textit{during} computation.  This is in contrast with nearly all current digital quantum algorithms, where a preset sequence of quantum gates are applied, and the system is measured only as the final step of the computation.  Next, Sec.~\ref{sec:qps} begins by briefly reviewing the QPS algorithm~\cite{qps}, including its qubit and quantum gate requirements.  Section~\ref{sec:qps} then continues by introducing a modified QPS that incorporates mid-circuit measurement and quantum-classical feedback.  We have implemented this algorithm in \texttt{Qiskit}~\cite{Qiskit} and provide numerical results in Sec.~\ref{sec:numerical}.  The simulations agree with those in Ref.~\cite{qps} and we are able to make more precise predictions than were possible before, using the matrix product state simulator. Our Python code is available at \href{https://github.com/LBNL-HEP-QIS/QPS_public}{https://github.com/LBNL-HEP-QIS/QPS\_public}. The paper ends with conclusions in Sec.~\ref{sec:conclusions}.

\section{Dynamic Quantum Computing}
\label{sec:dynamic}
Dynamic quantum computing involves dividing a program into (1) steps that can only be implemented on a quantum computer (QPU) and (2) steps that can be implemented more efficiently on a classical computer (CPU), and (3) interfacing between the QPU and CPU wherever necessary. 
This scheme manifests in two categorically distinct ways.

First, one could execute an algorithm consisting of a sequence of alternating quantum and classical steps, where the result of each step is fed serially to the next.
The variational quantum eigensolver (VQE) \cite{vqe} is the primary example of this scheme.
To compute the smallest eigenvalue of a Hermitian operator $\mathcal{H}$, VQE alternates between a quantum step that computes the expectation of $\mathcal{H}$ on some vector $\ket{\psi(\vec{\theta})}$, and a classical step that minimizes the expectation $\langle\psi(\vec{\theta})|\mathcal{H}|\psi(\vec{\theta})\rangle$ over parameters $\vec{\theta}$.
In this scheme, each quantum step is independent from the previous, so the same QPU is reset and re-used for all quantum steps.
This means that the QPU's coherence time must be long enough only to accommodate the execution time for a single quantum step.

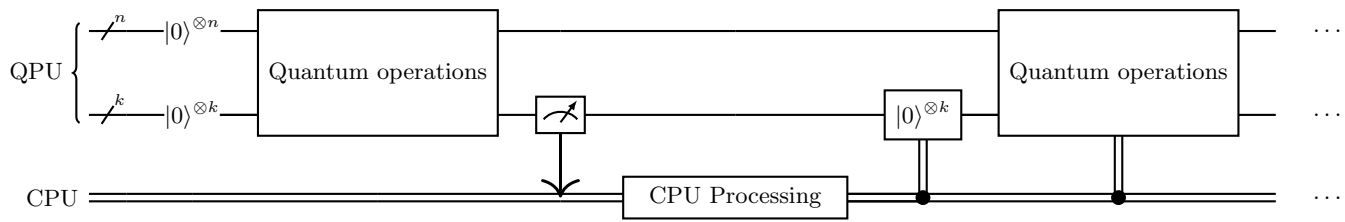
\begin{figure*}
 \begin{adjustbox}{width=1.0\textwidth}
    \begin{tikzcd}[]
        \lstick[wires=2]{QPU} &  \qwbundle{n} & \qw \ket{0}^{\otimes n} & \gate[2]{\mbox{$\text{Quantum operations}$}} & \qw & \qw & \qw &  \gate[2]{\mbox{$\text{Quantum operations}$}} & \qw & \dotsc\\
                              &  \qwbundle{k} & \qw \ket{0}^{\otimes k} & \qw & \meter{} \arrow[line width=0.04cm]{d} & \qw & \gate{\ket{0}^{\otimes k}} & \qw & \qw & \dotsc \\
        \lstick{CPU}                   &  \cw          & \cw                   & \cw & \cw & \gate[cwires={1}][3cm]{\mbox{$\text{CPU Processing}$}} & \cw \cwbend{-1} & \cwbend{-1} & \cw & \dotsc
    \end{tikzcd}
 \end{adjustbox}
    \caption{Dynamic computing workflow. The essential procedure consists of four steps: (1) Measure a subset of quantum resources in the QPU, represented here by the $k$-qubit register, (2) On the CPU perform some processing on the measured data, (3) Reset the measured qubits to the $\ket{0}$ state so they can be re-used, and (4) Based on CPU outputs, apply additional quantum operations on the QPU. Note that the QPU must maintain coherence throughout this procedure.}
    \label{fig:workflow}
\end{figure*}

On the other hand, one could also construct an algorithm that requires rapid interfacing between QPU and CPU.
For example, standard quantum error correction (QEC) procedures entail measuring a \textit{syndrome} operator, followed by a \textit{recovery operation} in which pre-defined quantum gates are applied conditional on measurement results \cite{nielsen_chuang_2010}.
Therefore, incorporating QEC to some general quantum program requires frequent interfacing between QPUs and CPUs in the form of measurements (QPU to CPU) and feedback (CPU to QPU).
In addition to direct feedback based on measurement results, one could also perform classical computations on them before applying quantum gates conditioned on the results of those computations.
This procedure describes a fully dynamic quantum-classical computer and is the basis for the algorithm presented in this paper.
Figure~\ref{fig:workflow} illustrates the operation of a dynamic quantum computer.

In contrast with VQE, where quantum and classical steps alternate serially, QEC and other algorithms that use a similar rapid interfacing scheme  require that quantum resources (qubits) maintain coherence during quantum-classical interfacing.
Limited qubit coherence times 
 are a major bottleneck for implementing QEC and other dynamic algorithms.
However, coherence times continue to improve as hardware is developed and refined, and because of the importance of QEC for developing fault-tolerant quantum computers, dynamic hardware will be a major focus in the long term.
Until now, demonstrations of dynamic computing with rapid digital interfacing include active qubit resets \cite{Riste2012, Salathe2018, Xiang2020}, quantum teleportation \cite{Steffen2013, Barrett2004, Riebe2004, Chou2018, Furusawa1998}, and error-correction \cite{Riste2020, Ofek2016, CampagneIbarcq2013, Minev2019, Pittman2005, Chiaverini2004, Vijay2012, Hu2019, Andersen2019, Bultink2020}.  A more complex example was recently demonstrated by IBM~\cite{dynamic_ibm}, employing a hybrid quantum-classical version of phase estimation on two qubits, where an $m$-bit representation of the phase is computed using several shots of a hybrid circuit that contains $m$ measurement-feed-forward cycles.
For each cycle, an $R_z(\theta)$ gate is selected and applied conditionally on previous measurement results.
Different measurement results select different $\theta$ values.

The hybrid phase estimation algorithm demonstrated by IBM \cite{dynamic_ibm} closely resembles the hybrid QPS we will introduce below in that it consists of single-qubit rotation gates with rotation angles that depend on classical information, i.e. mid-circuit measurement results.
Therefore, IBM's demonstration of executing a hybrid quantum algorithm shows that our QPS algorithm can theoretically be implemented on real devices in the future.
In \cref{sec:qps}, we summarize the original QPS algorithm \cite{qps} before laying out the dynamical version.

\section{Quantum Parton Shower Algorithm}
\label{sec:qps}
First, we summarize the quantum parton shower (QPS) algorithm originally presented in Ref.~\cite{qps}. Slightly modified variants of QPS are also presented in \cite{Bepari_2021, qps_random_walk}.

\vspace{-4pt}
\subsection{Physical Background}
\label{subsec:background}

Parton shower algorithms are perturbative approaches to efficiently describe high-multiplicity final states by focusing on the soft and collinear regions of phase space.  The QPS algorithm in Ref.~\cite{qps} was developed for a simple quantum field theory involving two types of fermion fields, $f_1$ and $f_2$, interacting with
one scalar boson $\phi$ governed by the following Lagrangian:
\begin{align}\nonumber
    \mathcal{L} \,=\, \bar{f}_1 & (i\slashed{\partial} + m_1)f_1 + \bar{f}_2(i\slashed{\partial} + m_2)f_2 + (\partial_{\mu}\phi)^2 \\
    & + g_1\bar{f}_1f_1\phi + g_2\bar{f}_2f_2\phi + g_{12}\left[\bar{f}_1f_2 + \bar{f}_2f_1\right]\phi
    \label{eq:Lagrangian}\,.
\end{align}
The first three terms in \cref{eq:Lagrangian} describe the kinematic properties
of the fermions and scalar while the latter three terms
govern their interactions.  The goal of a PS algorithm is to describe the collinear dynamics of the theory, which in this case correspond to the fermions radiating
scalars ($f_i\rightarrow f_j\phi$) and scalars spliting into fermion pairs ($\phi\rightarrow f_i\bar{f}_j$).

In classical PSs, the rates of these processes are described by splitting functions:
\begin{align}
    P_{i\rightarrow j\phi}(\theta) &= g_{ij}^2 P_f(\theta) \,,\quad i,j\in\{1,2\} \\
    P_{\phi\rightarrow i\bar{j}}(\theta) &= g_{ij}^2 \hat{P}_{\phi}(\theta) \,,\quad i,j\in\{1,2\}\,,
    \label{eq:split_fcns}
\end{align}
where $g_i\equiv g_{ii}$.  The splitting functions describe the probability for a particular particle at a given step (parameterized by the scale $\theta$) in the parton shower evolution to undergo a transformation.  There are many formally equivalent
definitions of the scale; here we use a common choice of the
opening angle of the emission with respect to the emitter (angularly ordered shower).

In addition to the splitting functions, another important
quantity is the no-branching probability (Sudakov factor):
\begin{align}
    \Delta_{i,k}(\theta_1, \theta_2) = \exp\left[-g_i^2\int_{\theta_1}^{\theta_2}d\theta'\hat{P}_k(\theta')\right]\,,
    \label{eq:sudakov}
\end{align}
which describes the probability that no emission occurs between scales $\theta_1$ and $\theta_2$. With the splitting functions and Sudakov factors, we can sample from the cross-section using a Markov Chain algorithm that generates one emission at a time, conditioned on the previous emissions.  In particular, at a given step in the algorithm with a fixed number of particles, the probability that none of them radiate or split is simply a product over Sudakov factors.
If something does happen at a given step, the probabilities are proportional to the appropriate splitting function.
In the limit $g_{12}\rightarrow0$, the Markov Chain algorithm can be implemented in terms of emission probabilities, and is therefore classically efficient.
However, if $g_{12}>0$, there are now multiple histories with
unmeasured intermediate fermion types which contribute to the same final state.
To account for these interferences, we must implement the Markov Chain at the quantum amplitude level -- and with the range of opening angles $\theta$ discretized into $N$ parts, this necessitates keeping track of $\bO(2^N)$ different histories.
This motivates the QPS algorithm, which computes the final state radiation with $g_{12}>0$ using only polynomially many qubits and gates on a digital quantum computer.

\subsection{Basis for the Quantum Algorithm}
\label{subsec:basis}
The interaction terms (\cref{eq:Lagrangian}) of the Lagrangian can be written as a matrix equation:
\begin{align}
    \mathcal{L}_{\text{interaction}}\,=\,\left(\begin{matrix}\bar{f}_1&\bar{f}_2\end{matrix}\right)\left(\begin{matrix}g_1&g_{12}\\g_{12}&g_2\end{matrix}\right)\left(\begin{matrix}f_1\\f_2\end{matrix}\right)\,\phi\,.
\end{align}
Furthermore, the ``interaction matrix'' is real and symmetric, and can thus be diagonalized as
\begin{align}
    G \,\equiv\, \left(\begin{matrix}g_1&g_{12}\\g_{12}&g_2\end{matrix}\right) \,=\, U^{\dagger}\left(\begin{matrix}g_a&0\\0&g_b\end{matrix}\right)U\,.
    \label{eq:diagonalization}
\end{align}
By defining a change of basis
\begin{align}
    \left(\begin{matrix}f_a\\f_b\end{matrix}\right) \,\equiv\, U\left(\begin{matrix}f_1\\f_2\end{matrix}\right)\,,
    \label{eq:basis_operation}
\end{align}
the interactions (\cref{eq:Lagrangian}) become diagonal:
\begin{align}
    \mathcal{L}_{\text{interaction}}\,=\,\left(\begin{matrix}\bar{f}_a&\bar{f}_b\end{matrix}\right)\left(\begin{matrix}g_a&0\\0&g_b\end{matrix}\right)\left(\begin{matrix}f_a\\f_b\end{matrix}\right)\,\phi\,.
\end{align}

In this ``diagonal basis'', splitting does not create interference between fermion types.
In other words,
\begin{align}
    P_{i\rightarrow j\phi}(\theta) &= \delta_{ij}g_{ij}^2 P_f(\theta) \,,\quad i,j\in\{a,b\} \\
    P_{\phi\rightarrow i\bar{j}}(\theta) &= \delta_{ij}g_{ij}^2 P_{\phi}(\theta) \,,\quad i,j\in\{a,b\}\,,
\end{align}
where $\delta_{ij}$ is the Kronecker delta. Note that this is also the case in the original basis if $g_{12}=0$.

Therefore, to simulate interference between fermion types, we first rotate particle registers $\ket{p}$ encoding fermion/boson fields into the diagonal basis according to \cref{eq:basis_operation}, proceed with a quantum analogue of the classical Markov Chain algorithm (generating a \textit{history} of angles and particle types), and lastly rotate the final particle states back to the original basis. 
If $g_{12}>0$, then the initial rotation to the diagonal basis creates a superposition of $f_a$ and $f_b$ fermions.
Subsequent operations act on this superposition, and all intermediate amplitudes/histories are preserved throughout the quantum Markov Chain.
This contrasts the classical MCMC parton shower, where superpositions of multiple fermion types are not included.

\subsection{Original Quantum Algorithm}
\label{subsec:quantum_algo}
The QPS circuit uses a qubit register each to encode the particle state and history, another three registers to store derived quantities about the number of particles, and a number of ancillary qubits.  For $N$ steps of the algorithm (i.e, discretizing the range of opening angles into $N$ parts) and with $n_I$ initial particles, the qubit counts for all of these registers are provided in \cref{tab:qps_registers}.
\begin{table}[htp]
    \centering
    \begin{tabular}{cll}
         Register & Purpose & \# qubits \\
         $\ket{p}$ & Particle state &   $3(N+n_I)$ \\
         $\ket{h}$ & Emission history & $\sum_{m=0}^{N-1}\ceil{\log_2(n_I+m+1)}$ \\
         $\ket{e}$ & Did emission happen? & $1$ \\
         $\ket{n_{\phi}}$ & Number of $\phi$ & $\ceil{\log_2(N+n_I)}$ \\
         $\ket{n_a}$ & Number of $f_a$ & $\ceil{\log_2(N+n_I)}$ \\
         $\ket{n_b}$ & Number of $f_b$ & $\ceil{\log_2(N+n_I)}$ \\
                     & Ancillas        & $4\ceil{\log_2(N+n_I)} + 5$ 
    \end{tabular}
    \caption{Registers in the QPS quantum circuit \protect{\cite{qps}} along with the number of qubits required for $N$ steps and $n_I$ initial particles.}
    \label{tab:qps_registers}
\end{table}

The QPS algorithm then proceeds by iteratively applying a series of steps.  As shown schematically in \cref{fig:qps_circ_orig}, each step $(m)$ has six components: (1) basis rotation $R^{(m)}$, (2) count particles $U_{\text{count}}$, (3) determine emission $U_e^{(m)}$, (4) update history $U_h$, (5) update particles $U_p^{(m)}$, and (6) inverse basis rotation $R^{(m)^{\dagger}}$.
Using the registers described in \cref{tab:qps_registers}, \cref{alg:1} gives a high-level description of the QPS algorithm in terms of these six meta gates.  
A detailed description of each register and each quantum gate is provided in \cref{app:1}.
\begin{algorithm}
    \DontPrintSemicolon
    \KwData{Splitting functions $P_{i\rightarrow j\phi}$, $P_{\phi\rightarrow ij}$, couplings $g_1$, $g_2$, $g_{12}$, step parameter $\epsilon$, number of steps $N$, and $n_I$ initial particles.}
    \KwResult{Full amplitude description of final state radiation.}
    \Begin{
        Initialize all qubit registers in the $\ket{0}$ state. \;
        Encode initial particles $\ket{p}_{1}$\, ... \, $\ket{p}_{n_I}$. \;
        \For{$j \gets 0$ \KwTo $N-1$}{
            (1) \textbf{Basis rotation:} Rotate all particles in $\ket{p}$ to the diagonal basis (\cref{eq:diagonalization}) using \cref{eq:basis_rotation}. \;
            (2) \textbf{Count particles:} Using $U_{\text{count}}$ (\cref{subsubsec:Ucount_old}), count the number of each particle type, storing the results in $\ket{n_a}$, $\ket{n_b}$, $\ket{n_{\phi}}$. \;
            (3) \textbf{Determine emission:} Using $U_e$ (\cref{subsubsec:Ue_old}), encode whether an emission occurred this step on $\ket{e}$, where the probability of emission is controlled on $\ket{n_a}$, $\ket{n_b}$, $\ket{n_{\phi}}$. \;
            (4) \textbf{Update history:} Using $U_h$ (\cref{subsubsec:Uh_old}), update the history register $\ket{h}_m$, which encodes \textit{which} particle (if any) emitted this step. The relative amplitudes for particular emissions are controlled by $\ket{n_a}$, $\ket{n_b}$, $\ket{n_{\phi}}, \ket{p}, \ket{e}$. (Note that $\ket{e}$ is put back into the $\ket{0}$ state implicitly in $U_h$.)\;
            (5) \textbf{Update particles:} Using $U_p$ (\cref{subsubsec:Up_old}), update the particle state, controlled on which particle emitted (encoded in $\ket{h_m}$). \;
            (6) \textbf{Inverse basis rotation:} Rotate all particles in $\ket{p}$ back to the original basis, using the inverse of \cref{eq:basis_rotation}.
        }
        Measure all qubits.
    }
    \caption{Original QPS algorithm \protect{\cite{qps}}}
    \label{alg:1}
\end{algorithm}
Additionally, \cref{fig:qps_circ_orig} illustrates two steps of the circuit diagram for \cref{alg:1}.
\begin{figure*}[htp]
 \begin{adjustbox}{width=0.97\textwidth}
    \begin{tikzcd}[column sep=0.14cm]
    \lstick{\ket{p}} & \qwbundle \qw & \gate{R^{(0)}} &   \measure{\mbox{$p$}} \vqw{4}  & \qw  & \measure{\mbox{$p$}} \vqw{2}  &  \gate{U^{(0)}_p} &  \gate{{R^{(0)}}^{\dagger}} & \slice{} \qw & \qw & \gate{R^{(1)}} &   \measure{\mbox{$p$}} \vqw{4}  & \qw  & \measure{\mbox{$p$}} \vqw{1}  &  \gate{U^{(1)}_p} &  \gate{{R^{(1)}}^{\dagger}}  & \qw & \dotsc   \\
    \lstick{$\ket{h}_1$} & \qwbundle \qw  & \qw & \qw & \qw & \qw  &   \qw  & \qw & \qw & \qw & \qw & \qw & \qw & \gate{U_{h}^{(1)}} \vqw{2}  &   \measure{\mbox{$h$}} \vqw{-1}  & \meter{} & & \\ 
    \lstick{$\ket{h}_0$} & \qwbundle \qw  & \qw & \qw & \qw & \gate{U_{h}^{(0)}} \vqw{1}  &   \measure{\mbox{$h$}} \vqw{-2}  & \meter{} & & & & & & & & & & \\ 
    \lstick{\ket{e}} & \qw  & \qw & \qw & \gate{U^{(0)}_{e}} &  \ctrl{1}   & \qw & \qw & \qw & \qw  & \qw & \qw & \gate{U^{(1)}_{e}} &  \ctrl{1}  & \qw & \qw  & \qw & \dotsc   \\
    \lstick{\ket{n_{\phi}}} & \qwbundle \qw & \qw  & \gate[3]{U_{\rm count}} & \measure{\mbox{$n_\phi$}} \vqw{-1}& \gate[3]{U_{h}} & \qw  & \qw  &\qw  & \qw & \qw  \qw  & \gate[3]{U_{\rm count}} & \measure{\mbox{$n_\phi$}} \vqw{-1}& \gate[3]{U_{h}} & \qw  & \qw  & \qw & \dotsc     \\
    \lstick{\ket{n_a}} & \qwbundle \qw  & \qw &  \ghost{U_{\rm count}} & \measure{\mbox{$n_a$}} \vqw{-1}  &  \ghost{U_{h}} &  \qw & \qw  & \qw  & \qw & \qw  & \ghost{U_{\rm count}} & \measure{\mbox{$n_a$}} \vqw{-1}  &  \ghost{U_{h}} &  \qw & \qw  & \qw & \dotsc \\
    \lstick{\ket{n_b}} & \qwbundle \qw  & \qw & \ghost{U_{\rm count}}  &\measure{\mbox{$n_b$}} \vqw{-1} & \ghost{U_{h}} & \qw & \qw & \qw  & \qw   & \qw   & \ghost{U_{\rm count}}  &\measure{\mbox{$n_b$}} \vqw{-1} & \ghost{U_{h}} & \qw & \qw   & \qw & \dotsc 
    \end{tikzcd}
 \end{adjustbox}
 \caption{High-level circuit diagram of the first two steps of the QPS algorithm. Round gates indicate control qubits.}
 \label{fig:qps_circ_orig}
\end{figure*}
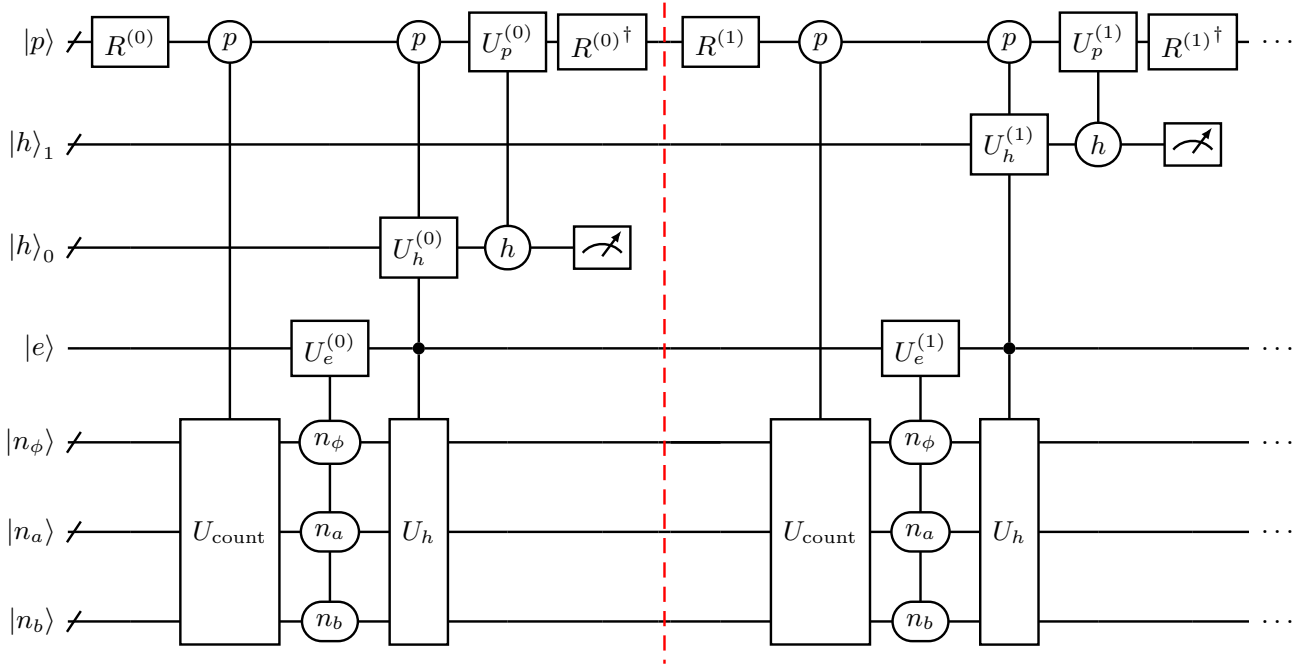
\begin{figure*}[htp]
 \begin{adjustbox}{width=0.97\textwidth}
    \begin{tikzcd}[column sep=0.2cm]
    \lstick{\ket{p}} & \qwbundle \qw & \gate{R^{(0)}} &   \measure{\mbox{$p$}} \vqw{3}  & \qw  & \measure{\mbox{$p$}} \vqw{3}  & \qw &  \gate{U^{(0)}_p} \vcw{1} & \gate{{R^{(0)}}^{\dagger}} & \slice{} \qw & \qw &  \gate{R^{(1)}} &   \measure{\mbox{$p$}} \vqw{3}  & \qw  & \measure{\mbox{$p$}} \vqw{1}  & \qw &  \gate{U^{(1)}_p} \vcw{1} & \gate{{R^{(1)}}^{\dagger}}  & \qw & \qw  & \dotsc   \\
    \lstick{\ket{h}} & \qwbundle \qw  & \qw & \qw & \qw & \gate{U_{h}^{(0)}}  &  \meter{}  & \qw \ket{0} & \qw & \qw & \qw & \qw & \qw & \qw & \gate{U_{h}^{(1)}} & \meter{} \qw & \qw \ket{0} & \qw & \qw & \qw  & \dotsc\\
    \lstick{\ket{e}} & \qw  & \qw & \qw & \gate{U^{(0)}_{e}} &  \ctrl{-1}   & \qw & \qw & \qw & \qw  & \qw & \qw & \qw & \gate{U^{(1)}_{e}} &  \ctrl{-1}  & \qw & \qw  & \qw & \qw & \qw & \dotsc   \\
    \lstick{\ket{n_{a}}} & \qwbundle \qw & \qw  & \gate{U_{\rm count}} & \measure{\mbox{$n_a$}} \vqw{-1}& \measure{\mbox{$n_a$}} \vqw{-1} & \qw  & \qw  & \qw & \qw & \qw & \qw  & \gate{U_{\rm count}} & \measure{\mbox{$n_a$}} \vqw{-1} & \measure{\mbox{$n_a$}} \vqw{-1} & \qw  & \qw  & \qw & \qw & \qw & \dotsc \\
    \lstick{$h_0$} & \cw & \cw & \cw & \cw & \cw & \gate[cwires={1}]{h_0} \vcw{-3} & \cwbend{-3} & \cw & \cw & \cw & \cw & \cw & \cw & \cw & \cw & \cwbend{-3} & \cw & \cw & \cw & \dotsc\\
    \lstick{$h_1$} & \cw & \cw & \cw & \cw & \cw & \cw & \cw & \cw & \cw & \cw & \cw & \cw & \cw & \cw & \gate[cwires={1}]{h_1} \vcw{-4} & \cwbend{-1} \vcw{-1} & \cw & \cw & \cw & \dotsc
    \end{tikzcd}
 \end{adjustbox}
 \caption{High-level circuit diagram of the first two steps of the improved QPS algorithm. Round gates indicate control qubits. Double wires indicate classical information stored on a CPU, and measured from $\ket{h}$. Double wires attached to $U_p$ indicate that sub-gates in $U_p$ are selectively applied based on measured values of $h_0,h_1,...$. The $\ket{0}$ attached to the same gate indicates a reset to the $\ket{0}$ state.}
 \label{fig:qps_circ_new}
\end{figure*}
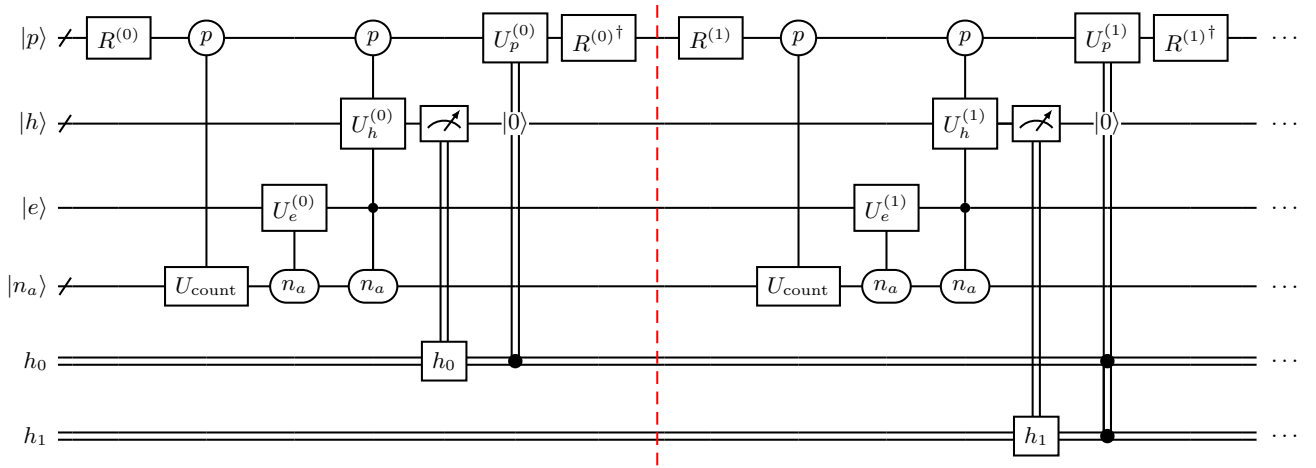
Note that emission history register $\ket{h}_m$ is updated and controlled on for $U_p$ once at step $m$, but untouched at previous and subsequent steps.  This property of the history register is a consequence of the lack of interference between different histories with the same emission angles.
This motivates the idea of measuring $\ket{h}_m$ after each step, and then re-using those qubits during subsequent steps, thus reducing qubit requirements for the QPS.

\subsection{Modified Quantum Parton Shower}
\label{subsec:modified_qps}
Using mid-circuit operations conditioned on measurement results, we can significantly simplify the quantum circuit of the previous section.
A qubit that is untouched for the remainder of a circuit can be measured at any time without changing (in principle) the properties of the encoded quantum state. In fact, a qubit used only to control operations on other qubits can be measured -- and all subsequent quantum controls replaced by classical controls -- without affecting the distribution of measured states. By invoking this deferred measurement principle on \cref{fig:qps_circ_orig}, the history sub-register $\ket{h}_m$ at step $m$ can be measured directly after applying $U_h^{(m)}$, and the results used to classically control $U_p^{(m)}$.
Therefore, in principle, the qubits of $\ket{h}_m$ can be reset to the ground state $\ket{0}$ and re-used for $\ket{h}_{m+1}$.
Because $\ket{h}_m$ encodes a superposition of \textit{which} particle $\ket{p_1},\dotsc,\ket{p_{n_I+m-1}},\,\text{or None}$ emitted at step $m$, measuring $\ket{h}_m$ projects the wavefunction onto a definite emission history.
In other words, the sequence of emission locations (particles) is stored classically during the circuit execution.
This does not affect the dynamics of the simulation, as different emission histories do not interfere with each other -- fermion superpositions within the particle register $\ket{p}$ remain intact and only affect operations within the same history. 
Therefore, by running the simulation a polynomial number of times (given some constant statistical tolerance), we still construct an accurate probability distribution of final states.
In this section, we describe at a high level how measuring and resetting $\ket{h}$ after each step simplifies the QPS algorithm.
A detailed treatment of the improvements is provided in \cref{app:1}.

Measuring the history register at a given step tells us \textit{which} particle emits at that step, so the entire emission history can stored as classical information during the simulation.
In fact, given that the simulation begins with  definite initial particle types (fermion or boson), then the \textit{type} of particle emitted at each step can be inductively inferred from the emission history.
Note that while individual fermions may be in a superposition of different ``flavors'' $a/b$ -- for the two types of emissions considered -- knowledge of the emitting particle type ($f$ or $\phi$) always implies the emitted particle type, as we consider the splittings $f\rightarrow f\phi$ and $\phi\rightarrow ff$.
Therefore, the total number of particles ($n_{\text{tot}}$), number of bosons ($n_{\phi}$), and number of fermions ($n_{f}=n_{f_a}+n_{f_b}$) are classical information.
This lets us remove two of the three counting registers $\ket{n_{\phi}}$, $\ket{n_a}$, $\ket{n_b}$, as follows.

In the original QPS, the $U_e$ and $U_h$ gates apply rotations with different angles, where each rotation (angle) is controlled on one of  the possible values stored in $\ket{n_a}\ket{n_b}\ket{n_{\phi}}$.
With mid-circuit measurements of $\ket{h}$, $n_{\phi}$ is classical information, so there is no need store and control on $\ket{n_{\phi}}$ at all. 
Now suppose we count the number of $a$-fermions using a quantum circuit -- without loss of generality we could instead count the number of $b$-fermions -- and apply rotations controlled on each possible value stored in $\ket{n_a}$. 
There is a one-to-one mapping of the possible values stored in $\ket{n_a}$ to the possible values of $\ket{n_b}$, given by $n_{\text{tot}} = n_a + n_b + n_{\phi}$, as $n_{\text{tot}}$ and $n_{\phi}$ are stored on a CPU.
Therefore, there is no need to store and control on $\ket{n_b}$, as the superposition of possible $n_b$'s is already implicitly encoded in $\ket{n_a}$.
Thus, by measuring $\ket{h}$ at each step, $\ket{n_b}$ and $\ket{n_{\phi}}$ become redundant and can be removed.

Suppose we start with 
\begin{align}
 \begin{split}
    n_{f,0} &= \text{initial number of fermions} = n_{f_a} + n_{f_b} \\
    n_{\phi,0} &= \text{initial number of bosons}\\
    n_{\text{tot},0} &= n_{f,0} + n_{\phi,0}
 \end{split}
\end{align}
Then, after measuring the history register, the CPU stores which of the initial particles emitted, so we adjust
\begin{align}
 \begin{split}
    n_{f,0} \,&\mapsto\, n_{f,1}\\
    n_{\phi,0} \,&\mapsto\, n_{\phi,1}\\
    n_{\text{tot},0} \,&\mapsto\, n_{\text{tot},1}
 \end{split}
\end{align}
in the CPU accordingly, and the emitted particle is encoded in $\ket{p_1}$.
This process is repeated for each subsequent simulation step, so inductively $n_{f,m}, n_{\phi,m}, n_{\text{tot},m}$ are stored in the CPU throughout the simulation.
Under the assumption that we can implement the workflow from \cref{fig:workflow}, we can use this information to reduce both the computational complexity and absolute qubit/gate counts of QPS.
\cref{alg:2} gives a high-level description of the improved QPS algorithm with mid-circuit measurements.
\begin{algorithm}
    \DontPrintSemicolon
    \KwData{Splitting functions $P_{i\rightarrow j\phi}$, $P_{\phi\rightarrow ij}$, couplings $g_1$, $g_2$, $g_{12}$, step parameter $\epsilon$, number of steps $N$, and $n_I$ initial particles.}
    \KwResult{Full amplitude description of final state radiation.}
    \Begin{
        Initialize all qubit registers in the $\ket{0}$ state. \;
        Encode initial particles $\ket{p}_{1}$\, ... \, $\ket{p}_{n_I}$. \;
        \For{$j \gets 0$ \KwTo $N-1$}{
            (1) Rotate all particles in $\ket{p}$ to the diagonal basis (\cref{eq:diagonalization}) using \cref{eq:basis_rotation}. \;
            (2) Using $U_{\text{count}}$ (\cref{subsubsec:Ucount_old}), count the number of $a$-fermions, storing the result in $\ket{n_a}$. \;
            (3) Using $U_e$ (\cref{subsubsec:Ue_old}), encode whether an emission occurred this step on $\ket{e}$, where the probability of emission is controlled on $\ket{n_a}$, and $n_{\phi}, n_{\text{tot.}}$ from the CPU. \;
            (4) Using $U_h$ (\cref{subsubsec:Uh_old}), update the history register $\ket{h}_m$, which encodes \textit{which} particle (if any) emitted this step. The relative amplitudes for particular emissions are controlled by $\ket{n_a}$, $\ket{e}$, and $n_{\phi}, n_{\text{tot.}}$ from the CPU. (Note that $\ket{e}$ is put back into the $\ket{0}$ state implicitly in $U_h$.)\;
            (5) Measure the history register $\ket{h}$, storing the result on the CPU.\;
            (6) If the measurement result indicates that a $\phi\rightarrow f\bar{f}$ emission occurred, apply $U_p$ (\cref{fig:Up_new}) to update the particle state. \;
            (7) Reset $\ket{h}$ to the $\ket{0}$ state. \;
            (8) Rotate all particles in $\ket{p}$ back to the original basis, using the inverse of \cref{eq:basis_rotation}.
        }
        Measure $\ket{p}$.
    }
    \caption{QPS with mid-circuit measurements}
    \label{alg:2}
\end{algorithm}

The second simplification is as follows.
Storing $\phi$ or None in particle register is redundant, as given the emission history, the location of all $\phi$ and None particles in the simulation is stored in the CPU.
Therefore, we encode particle $m$ into a qubit sub-register $\ket{p_m}$ only if it is a fermion.
To see that this can be done without significantly changing the QPS algorithm, we briefly comment on each gate that acts on $\ket{p}$: $R$, $U_{\text{count}}$, $U_h$, and $U_p$ (see \cref{fig:qps_circ_new}).
The basis rotation $R$ (\cref{eq:basis_rotation}) acts as the identity on $\phi$'s, so simply not encoding $\phi$'s does not affect the action of $R$.
For each particle $\ket{p_m}$ in $\ket{p}$, the counting gate $U_{\text{count}}$ applies an incrementer on $\ket{n_a}$ controlled on $\ket{p_m=f_a}$ (see \cref{subsubsec:Ucount_new}).
Therefore, not encoding $\phi$'s on qubits does not affect the action of $U_{\text{count}}$.
The $U_h$ gate applies a two-level rotation to $\ket{h}$ for each particle in $\ket{p}$, where the rotation angle is controlled by particle type (see \cref{subsubsec:Uh_new}).
Recall that by measuring the history register, the CPU stores whether each particle in the emission history is a $f$, $\phi$ or None.
Therefore, for fermions, control on $\ket{p}$ as before, but for $\phi$'s the corresponding rotation can simply be applied without controlling on $\ket{p}$.
In the circuit diagram of $U_h$ -- \cref{fig:Uh_new} -- this can be visualized by replacing quantum register $\ket{p_i}$ with classical wires if $\ket{p_i}=\ket{\phi}$.
The gate sequence of \cref{fig:Uh_new} is unchanged, except that some rotations $U_h^{(m,i)}$ may no longer be controlled on $\ket{p_i}$, but applied directly to $\ket{h}$.

Finally, the particle update $U_p$ must be applied only if a $\phi\rightarrow f\bar{f}$ emission occurs.
If that is the case, then the circuit in \cref{fig:Up_new} is applied to $\ket{p}$.

\subsubsection{Qubit Costs}
\label{subsubsec:qubit_costs}
Figure.~\ref{fig:qps_circ_new} illustrates the high-level circuit diagram for two steps of the improved QPS algorithm.
Note the reduction in registers compared to the original QPS.
The improved QPS circuit calls for four qubit registers, detailed in \cref{tab:qps2_registers}.
\begin{table}[htp]
    \centering
    \begin{tabular}{cll}
         Register & Purpose & \# qubits \\
         $\ket{p}$ & Particle state &   $2(N+n_I)$ \\
         $\ket{h}$ & Emission history & $\ceil{\log_2(N+n_I)}$ \\
         $\ket{e}$ & Did emission happen? & $1$ \\
         $\ket{n_a}$ & Number of $f_a$ & $\ceil{\log_2(N+n_I)}$ \\
                     & Ancillas        & $2\ceil{\log_2(N+n_I)} + 1$ 
    \end{tabular}
    \caption{Registers in the improved QPS algorithm along with the number of qubits required for $N$ steps and $n_I$ initial particles.}
    \label{tab:qps2_registers}
\end{table}

\begin{figure}[htp]
    \centering
    \includegraphics[width=\columnwidth]{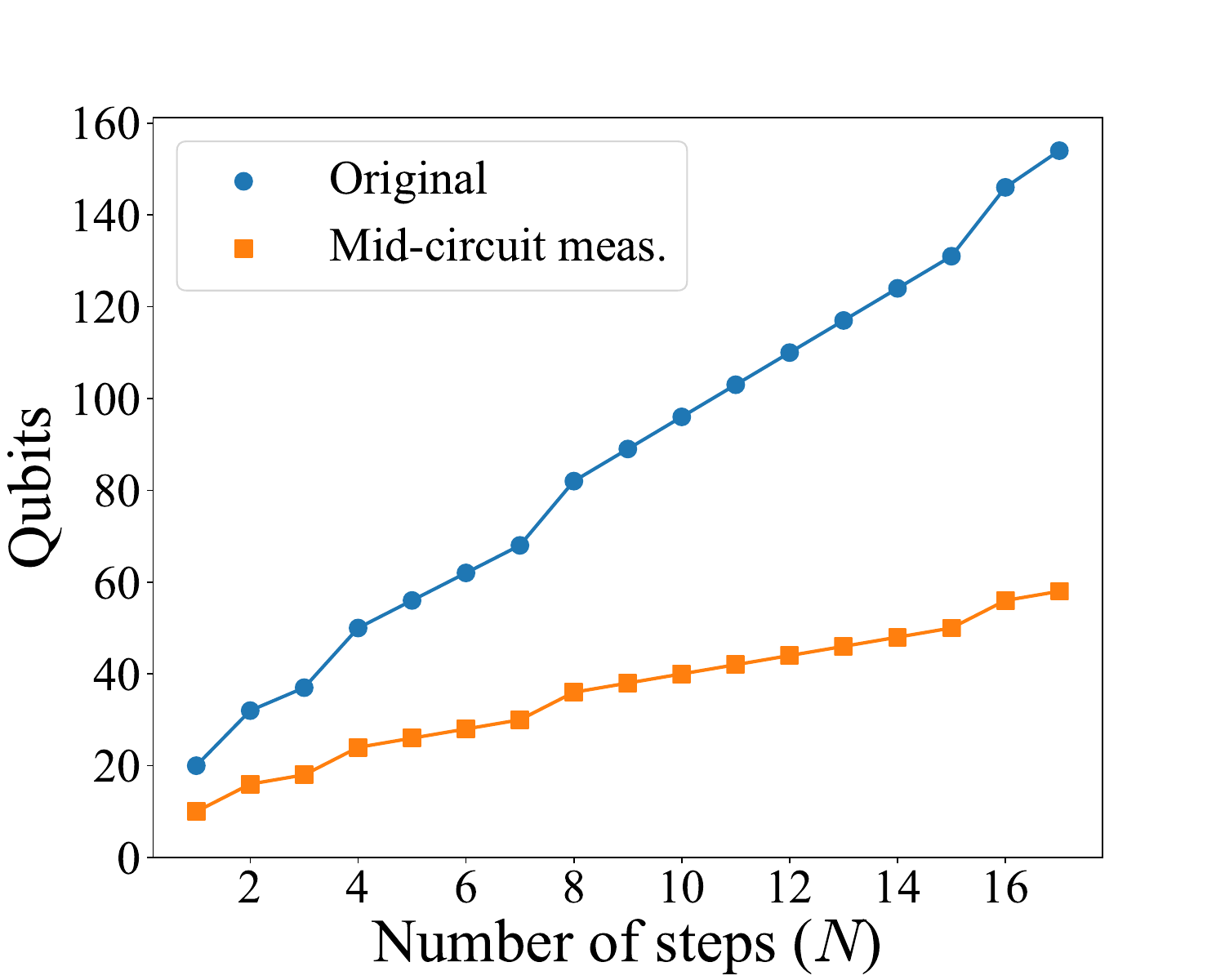}
    \caption{Qubit cost comparison between the original QPS and the improved version with mid-circuit measurements. This plot is simply an illustration of \protect{\cref{eq:num_qubits_new}} and the sum over Table \protect{\ref{tab:qps_registers}} with $n_I=1$. For $n_I>1$, the qubit count curves simply shift to the left, as $n_I$ and $N$ only appear together in the qubit scaling, as $N+n_I$.}
    \label{fig:qubit_costs}
\end{figure}

Note that the particle state only calls for $2(N+n_I)$ qubits, compared to $3(N+n_I)$ originally (Table.~\ref{tab:qps_registers}).
By only encoding fermions in quantum registers, two qubits are sufficient to encode $f_a$, $f_b$, $\bar{f}_a$, $\bar{f}_b$.
Also note that the number of required qubits varies between different circuit executions, as a simulation where more fermions are produced requires more quantum resources.
In the worst case, $\ket{p}$ will still consist of $N+n_I$ sub-registers, while the actual number is the total number of fermions in the system, denoted $n_f$.
Depending on parameters $g_1$, $g_2$, $g_{12}$, $\epsilon$, and the splitting functions $P_{i\rightarrow j\phi}$, $P_{\phi\rightarrow ij}$, the expectation for $\langle n_f\rangle$ may be significantly smaller than $N+n_I$.  
Therefore, the maximum number of qubits required for an $N$-step simulation is
\begin{align}
 \begin{split}
    \#\,\text{Qubit}&\text{s}_{\text{max}}\,=\\
    &\, 2(n_I+N) + 4\ceil{\log_2(n_I+N)} + 2 \,,
    \label{eq:num_qubits_new}
 \end{split}
\end{align}
while the actual number is 
\begin{align}
 \begin{split}
    \#\,&\text{Qubits}_\text{actual} \,=\\
            &2\Big[n_f + \ceil{\log_2(n_I+N)} + \ceil{\log_2(n_f)} + 1\Big] \,.
    \label{eq:num_qubits_new_exp}
 \end{split}
\end{align}
If a variable number of qubits can be used for circuit execution, the latter represents a significant resource reduction.
The asymptotic qubit scaling is just $\bO(N+n_I)$, compared to $\bO(N\log_2(N+n_I))$ for the original QPS algorithm (\cref{tab:qps_registers}).
This original qubit scaling is from storing the emission history at each step, which entails using $N$ subregisters $\ket{h}_m$ each with $\ceil{\log_2(n_I+m+1)}\sim\bO(\log_2(N+n_I))$ qubits.
By measuring, resetting, and re-using $\ket{h}$ after each step, $\ket{h}$ just consists of $\ceil{\log_2(N+n_I+1)}$ qubits, and the dominant scaling becomes the $\bO(N+n_I)$ qubits of $\ket{p}$.
The qubit scaling difference is illustrated in \cref{fig:qubit_costs}, which plots qubit count against $N$, with one starting particle, $n_I=1$.

\subsubsection{Gate Costs}
\label{subsubsec:gate_costs}
We measure gate costs by writing operations in terms of the universal standard gate set consisting of two-qubit controlled not gates (CNOTs) and arbitrary single qubit gates $U(\theta, \phi, \lambda)$.
Multi-controlled gates are decomposed into sequences of Toffoli (CCX) gates using a standard procedure that requires ancillary qubits equal to the number of controls minus one \cite{nielsen_chuang_2010}.
Then, Toffoli gates are decomposed into six CNOTs \cite{nielsen_chuang_2010}.
For an $n$-control NOT gate, this decomposition uses $\mathcal{O}(n)$ CNOTs.
As two-qubit entangling gates are far costlier to implement in real devices than single qubit gates, we quote gate counts in terms of CNOTs.
Note that while we illustrate ``classical controls'' in our circuit diagrams (\cref{fig:Up_new}, \cref{fig:Uh_new}) for quantum gates selected dynamically by the CPU, just the attached quantum gates are included in the gate count.

\cref{tab:gate_costs} summarizes the gate costs of each component of the improved QPS algorithm.
\begin{table}[htp]
    \centering
    \begin{tabular}{l|l|l}
         Element            & Step $m$ cost   &  Total Scaling ($N$ steps)   \\
         &&\\
         $R$                & $2(n_I+m)$           & $\bO(N^2+Nn_I)$        \\
         $U_{\text{count}}$ & $13\ell(n_I+m)$ & $\bO(N(N+n_I)\cdot\log_2(n_I+N))$\\
         $U_e$              & $(n_I+m)(12\ell-10)$ & $\bO(N(N+n_I)\cdot\log_2(n_I+N))$\\
         $U_h$              &                      & $\bO(N(N+n_I)^2\cdot\log_2(n_I+N)^2)$  \\
         $U_p$              & $2$ & $2N$ \\
         &&\\
         Total & & $\bO(N(N+n_I)^2\cdot\log_2(n_I+N)^2)$ 
    \end{tabular}
    \caption{Gate costs of the different circuit elements using re-measurement. $\ell\equiv\ceil{\log_2(n_I+m+1)}$}.
    \label{tab:gate_costs}
\end{table}
The overall asymptotic scaling is
\begin{equation}
    \bO(N\cdot(N+n_I)^2\cdot\log_2(N+n_I)^2) \,,
    \label{eq:new_scaling}
\end{equation}
which is a factor of $(N+n_I)^2$ more efficient than the original QPS gate scaling:
\begin{equation}
    \bO(N\cdot(N+n_I)^4\cdot\log_2(N+n_I)^2) \,.
    \label{eq:old_scaling}
\end{equation}
This scaling improvement is due to the fact that at step $m$, $\ket{n_a}$ is a superposition of $n_I+m$ possible basis states, while $\ket{n_a}\ket{n_b}\ket{n_{\phi}}$ is a superposition of $(n_I+m)^3$ possible basis states.
To implement $U_h$ (see \cref{app:1}), rotation gates controlled on the counting registers $\ket{n_i}$ are applied to $\ket{h}$ for each possible value stored in $\ket{n_i}$.
Therefore, in the original algorithm, $U_h$ consisted of $\bO((n_I+m)^3)$ controlled-rotations, while in the improved algorithm, only $\bO(n_I+m)$ controlled rotations are applied.

\begin{figure}[htp]
    \centering
    \includegraphics[width=\columnwidth]{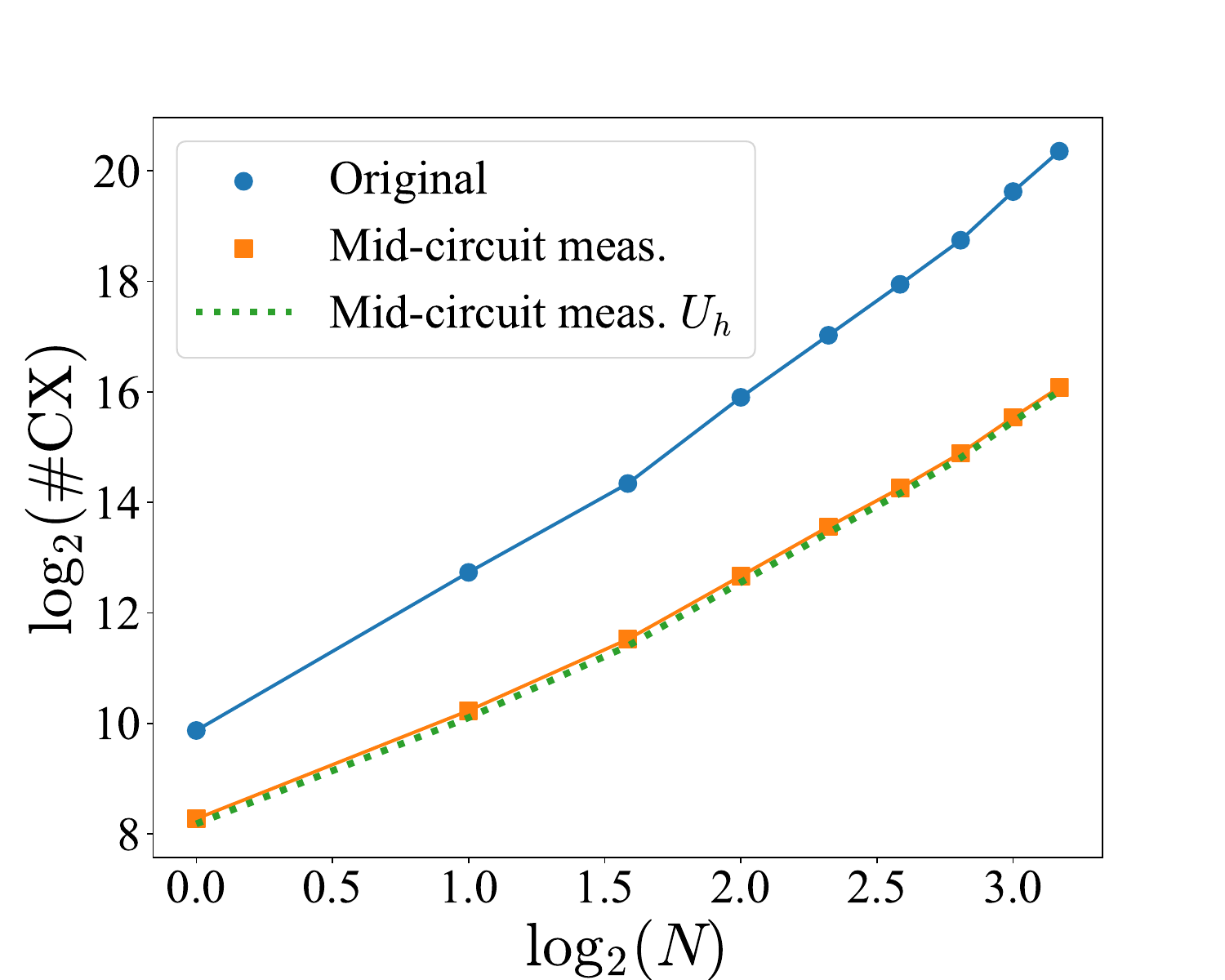}
    \caption{Gate cost comparison: The dashed line represents the dominant contribution from $U_h$ to the total gate count for QPS with mid-circuit measurements.}
    \label{fig:gate_costs}
\end{figure}

\cref{fig:gate_costs} compares the actual gate counts of our improved QPS circuits with those of the original QPS circuits.
The dashed line is the contribution from just the $U_h$ gate, and \cref{fig:gate_costs} illustrates that $U_h$ dominantes the gate count.

\section{Numerical Results}
\label{sec:numerical}
Using \texttt{Qiskit}'s matrix product state simulator~\cite{mps1,mps2}, we are able to simulate QPS with one initial particle ($n_I=1$) up to several steps.
For each simulation, we use a scale parameter $\theta=\theta_m$, defined by
\begin{align}
    \theta_m \equiv \epsilon^{m/N}\,, \quad \epsilon=0.001\,,
    \label{eq:params1}
\end{align}
splitting functions
\begin{align}
    P_{i\rightarrow i\phi}(\theta) &\,=\, g_i^2\hat{P}_f(\theta)\,=\, \frac{g_i^2\log(\theta)}{4\pi} \\
    P_{\phi\rightarrow i\bar{i}}(\theta) &\,=\, g_i^2\hat{P}_{\phi}(\theta)\,=\, \frac{g_i^2\log(\theta)}{4\pi}\,,
    \label{eq:splittings}
\end{align}
couplings
\begin{align}
    g_1= 2\, \qquad g_2 = 1 \, \qquad g_{12} = 1 \,,
    \label{eq:num_couplings}
\end{align}
and one initial $f_1$ (see \cref{eq:p_state}),
\begin{align}
    \ket{p}_{\text{initial}} = \ket{p_1} = \ket{100} \,.
\end{align}
The couplings $g_1=2, g_2=1, g_{12}=1$ are arbitrary, but chosen such that $g_a, g_b\neq 0$ and $g_a\neq g_b$ (\cref{eq:diagonalization}), in order to capture the full problem complexity.  For simplicity, the couplings are also kept independent of step. (In reality, they would run with the scale).  This means that the rotations are the same at each step.
The numerical values of the diagonalized couplings are
\begin{align}
    g_a \,=\, \frac{3+\sqrt{5}}{2} \,\approx\, 2.618\\
    g_b \,=\, \frac{3-\sqrt{5}}{2} \,\approx\, 0.382\,,
\end{align}
and the rotation angle $u$ is
\begin{align}
    u= \frac{\sqrt{5}-1}{2\sqrt{5}}\,\approx\,0.28\,\text{rad}\,.
\end{align}
With these parameters, the Sudakov factors, which give probabilities of no emission from a particular particle at step $m$, can be written as
\begin{align}
    \Delta_a(\theta_m) &\,=\, \epsilon^{g_a^2 / 4\pi N} \\
    \Delta_b(\theta_m) &\,=\, \epsilon^{g_b^2 / 4\pi N} \\
    \Delta_{\phi}(\theta_m) &\,=\, \epsilon^{(g_a^2+g_b^2) / 4\pi N} \,.
    \label{eq:sudakov_numerical}
\end{align}
Because the couplings are kept constant, these probabilities also remain constant at each step.

We run simulations with $g_{12}=0$ in addition to $g_{12}=1$.
As explained in \cref{subsec:background}, when $g_{12}=0$ the parton shower can be solved using a classical Markov Chain algorithm.
Therefore, as a sanity check, we overlay analytical Markov Chain calculations, each with $10^7$ shots, over simulation results with $g_{12}=0$ in our plots.

\begin{figure*}[htp]
    \captionsetup{font={Large}}
    \begin{subfigure}[t]{0.48\textwidth}
        \caption{$N=2$}
        \includegraphics[width=0.8\textwidth]{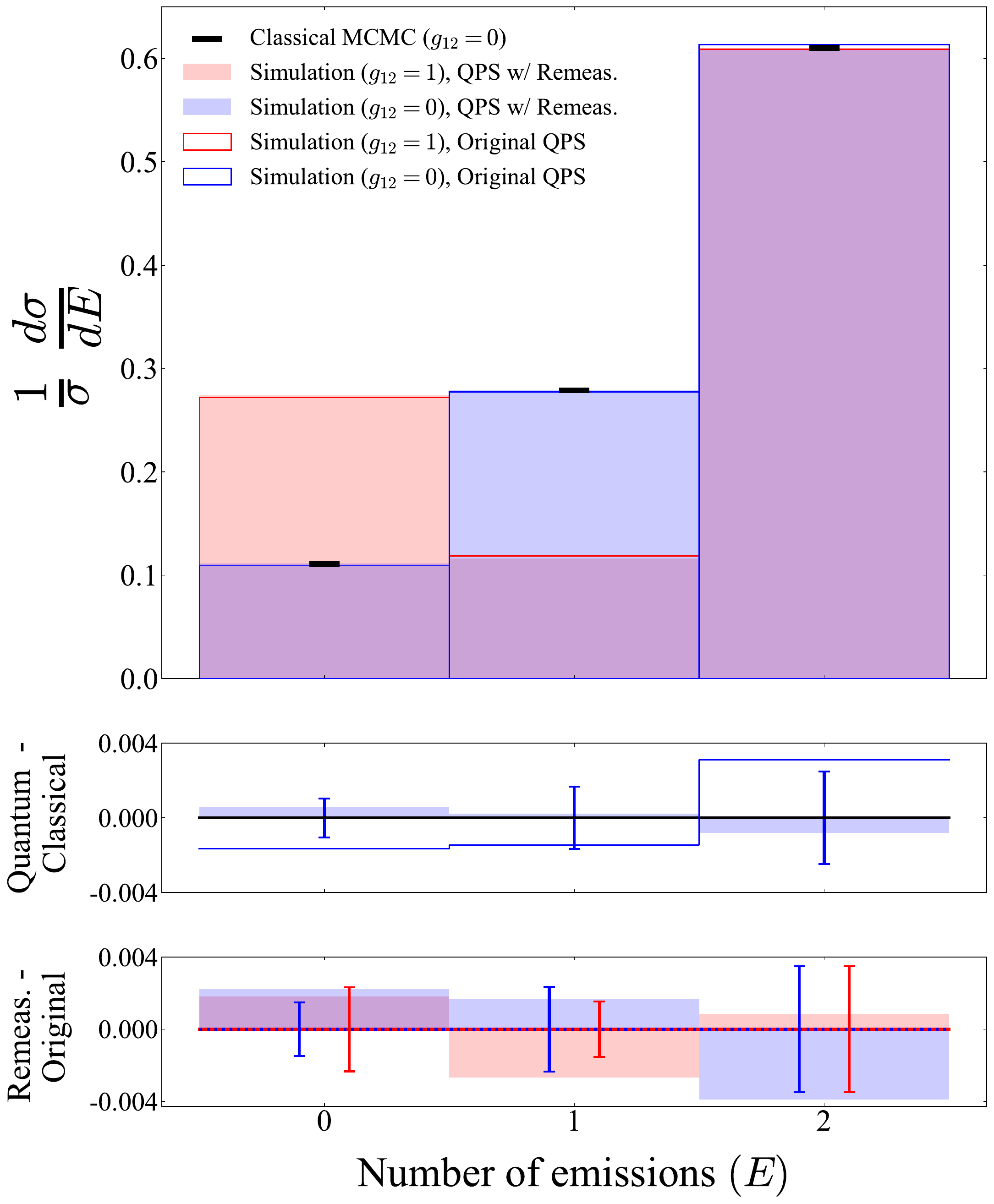}
    \end{subfigure}
    \begin{subfigure}[t]{0.48\textwidth}
    \caption{$N=3$}
        \includegraphics[width=0.8\textwidth]{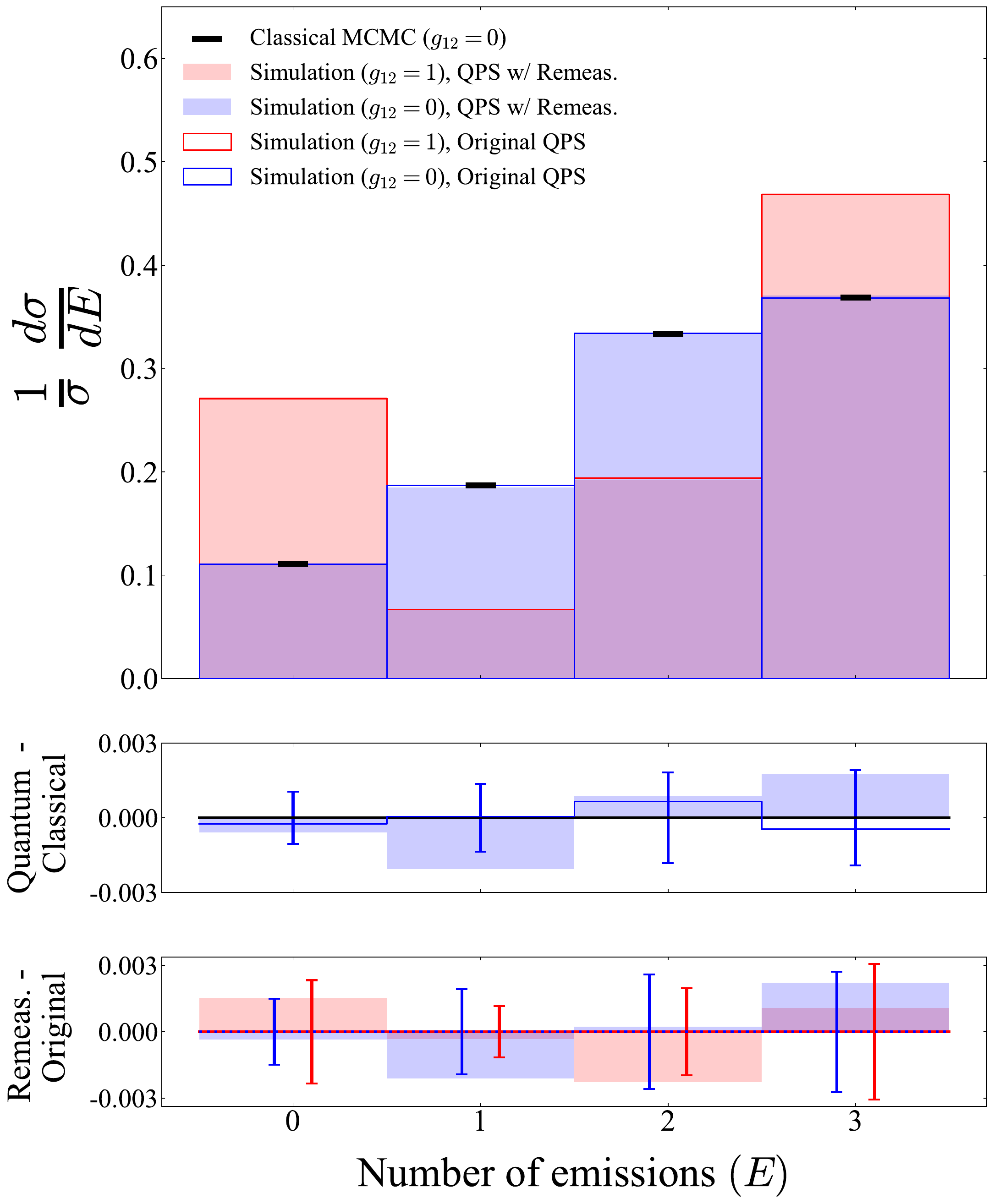}
    \end{subfigure}
    \vspace{14pt}\\
    \begin{subfigure}[t]{0.48\textwidth}
    \caption{$N=4$}
        \includegraphics[width=0.8\textwidth]{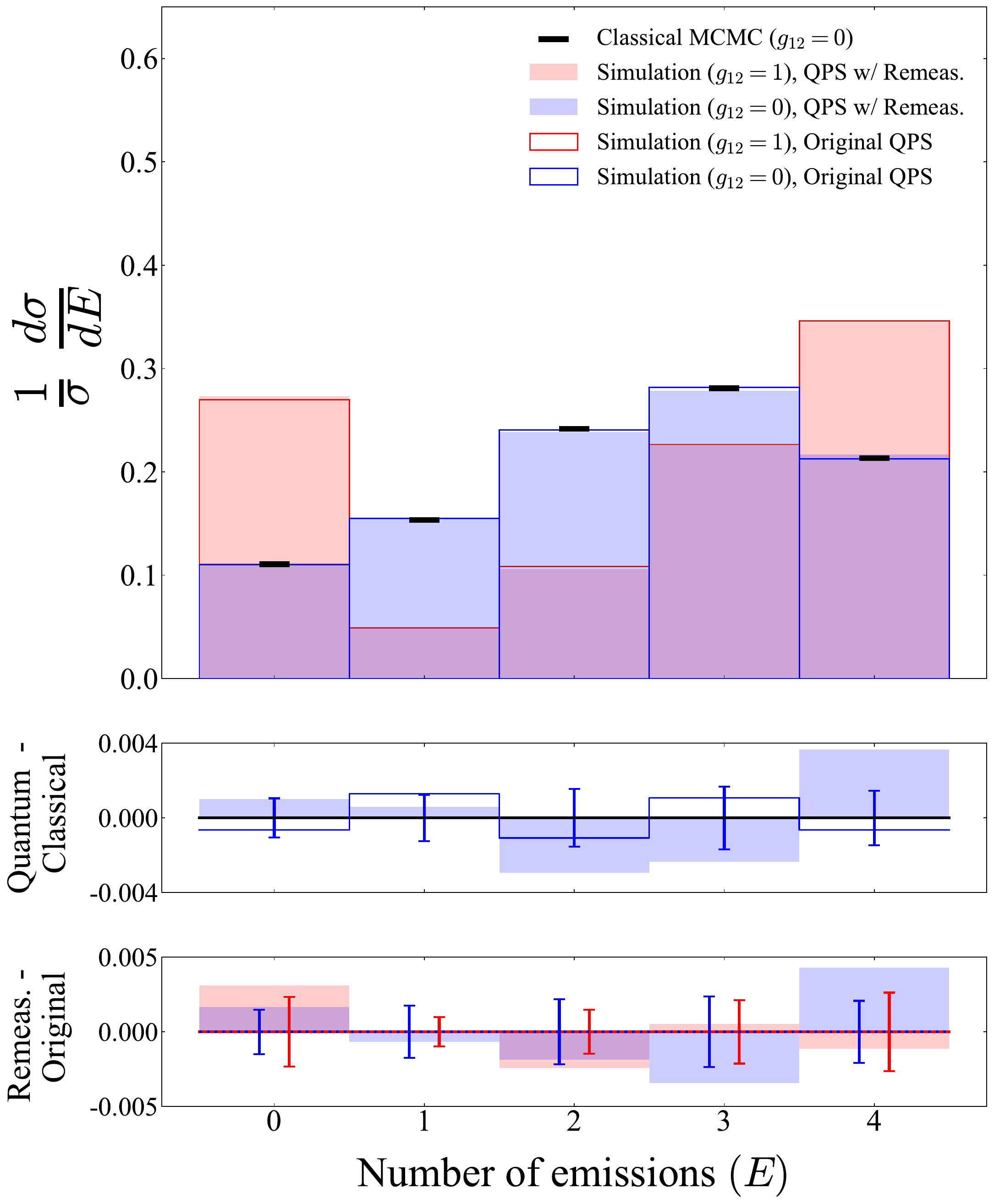}
    \end{subfigure}
    \begin{subfigure}[t]{0.48\textwidth}
    \caption{$N=5$}
        \includegraphics[width=0.8\textwidth]{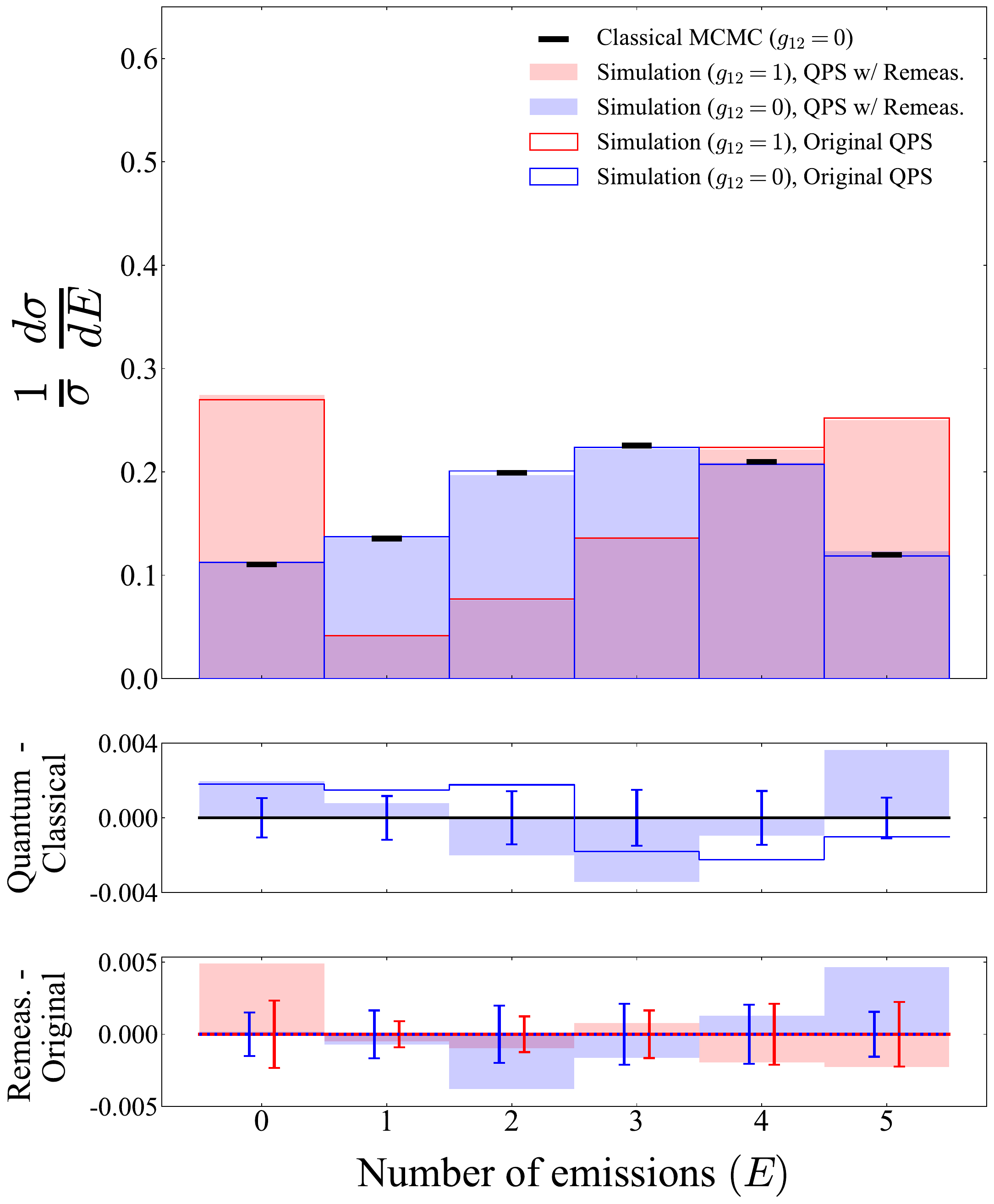}
    \end{subfigure}
    \captionsetup{font={normalsize}}
    \caption{Probability vs. Number of emissions ($E$) for 2, 3, 4, and 5 step simulations. Error bars represent $1\sigma$ ranges, e.g., in each third subplot, the red error bars correspond to the standard deviation of the difference distribution between simulation results obtained from original QPS and QPS with remeasurement. Classical MCMC data were obtained using $10^7$ shots, so the statistical errors are suppressed by a factor of $1/10$, and are thus negligible. Error bars in each second subplot are the statistical deviations for $g_{12}=0$ simulations.}
    \label{fig:N_emissions}
\end{figure*}

\begin{figure*}[htp]
    \captionsetup{font={Large}}
    \begin{subfigure}[t]{0.48\textwidth}
        \caption{$N=2$}
        \includegraphics[width=0.8\textwidth]{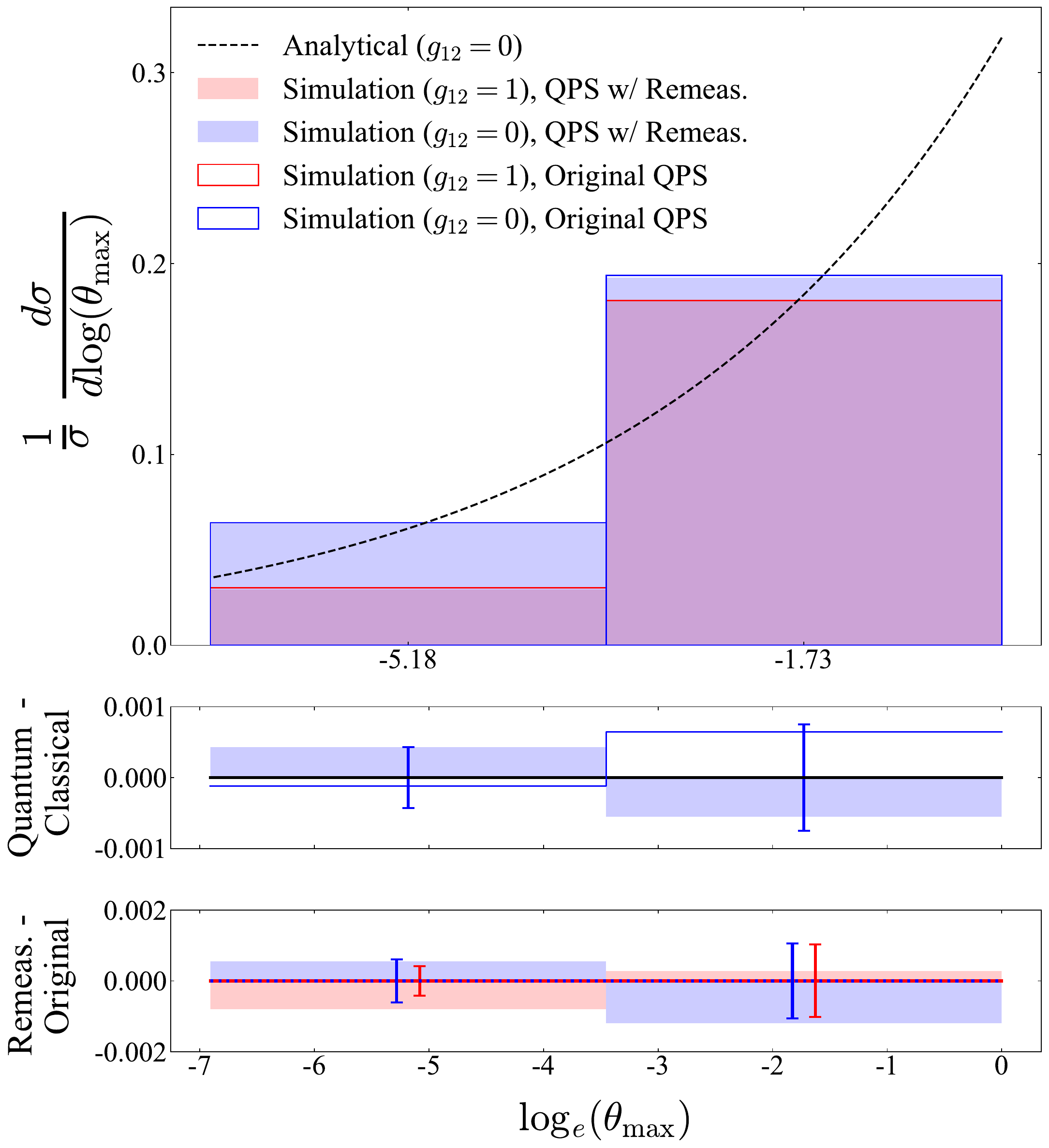}
    \end{subfigure}
    \begin{subfigure}[t]{0.48\textwidth}
    \caption{$N=3$}
        \includegraphics[width=0.8\textwidth]{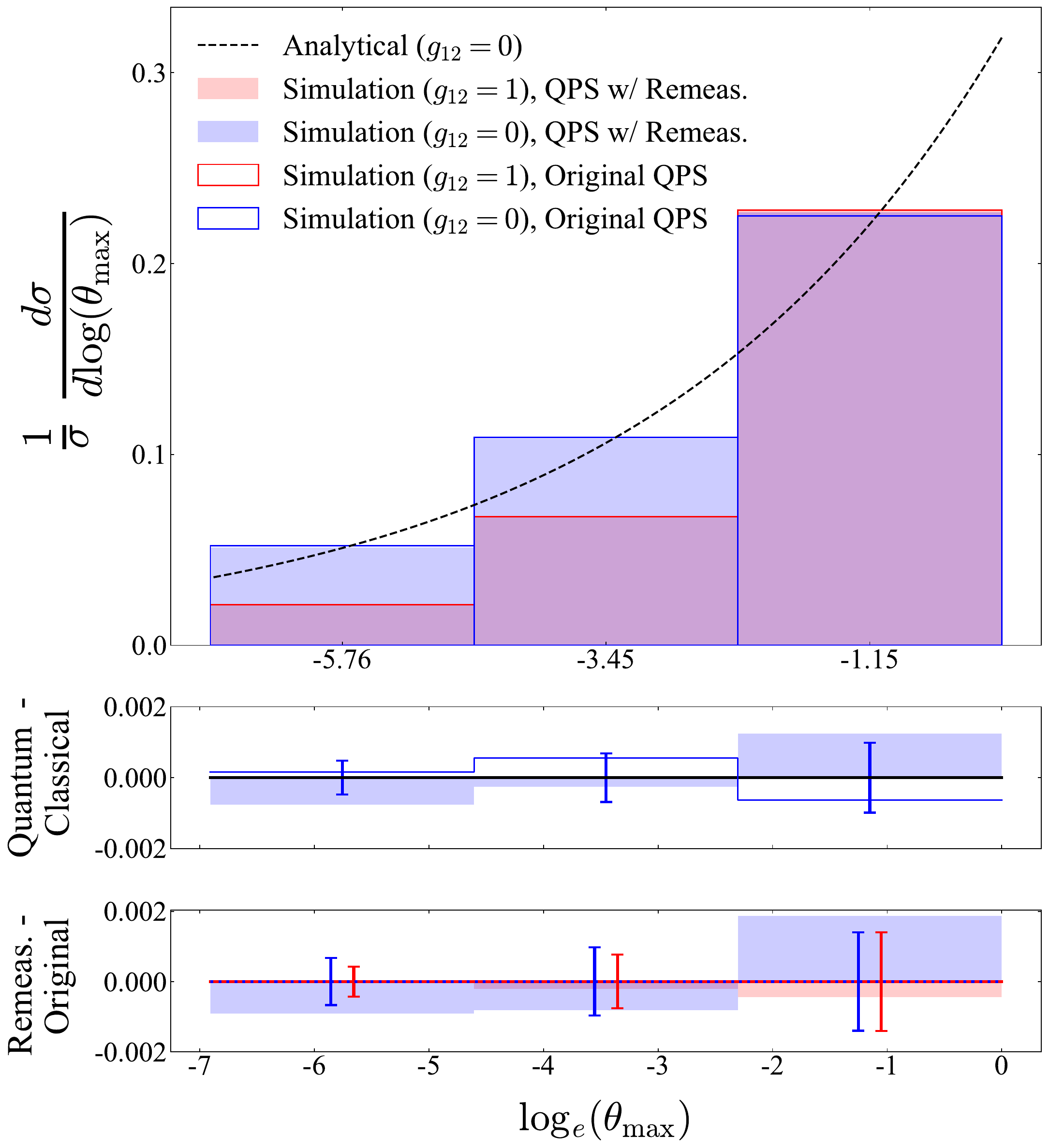}
    \end{subfigure}
    \vspace{14pt}\\
    \begin{subfigure}[t]{0.48\textwidth}
    \caption{$N=4$}
        \includegraphics[width=0.8\textwidth]{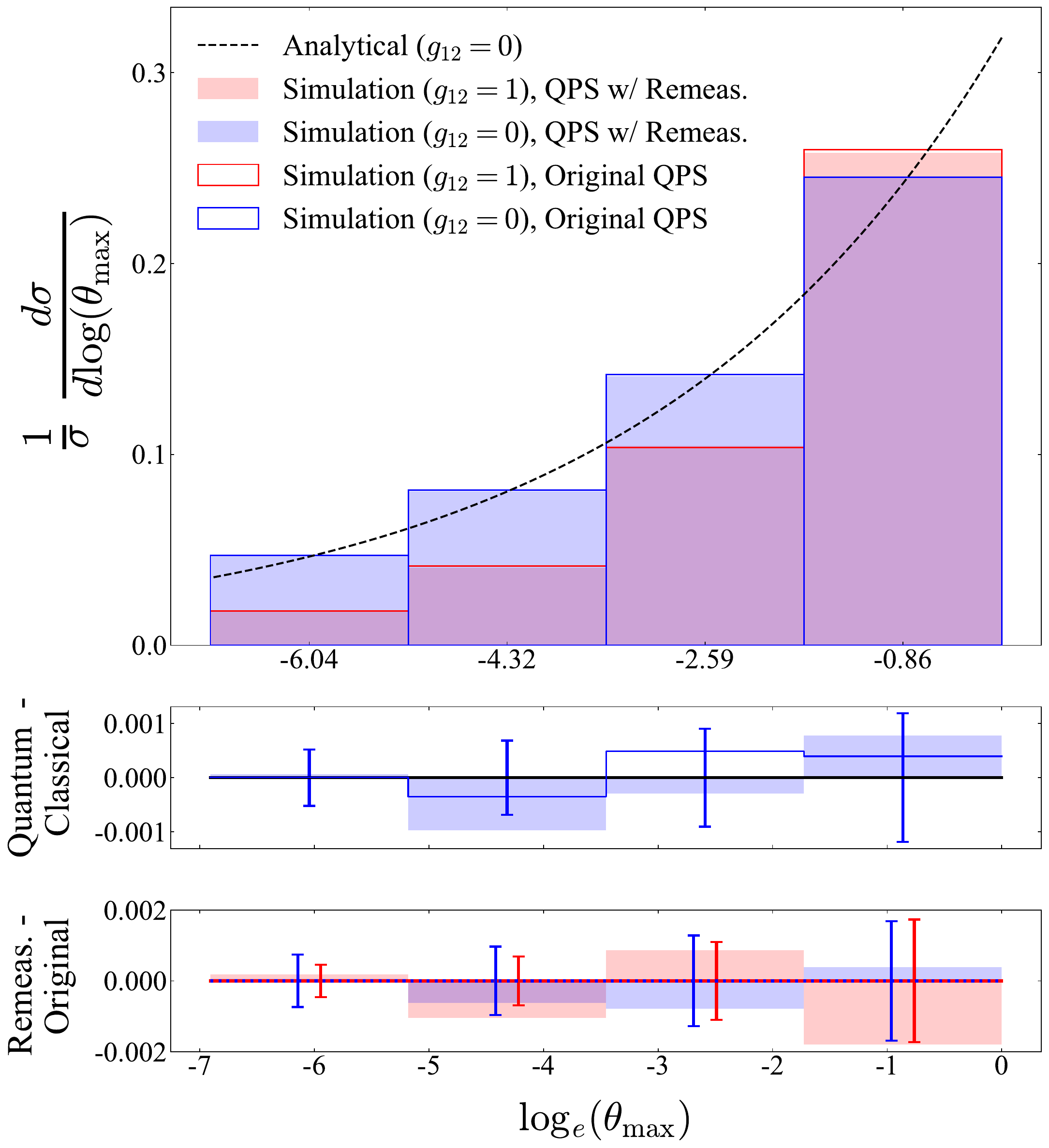}
    \end{subfigure}
    \begin{subfigure}[t]{0.48\textwidth}
    \caption{$N=5$}
        \includegraphics[width=0.8\textwidth]{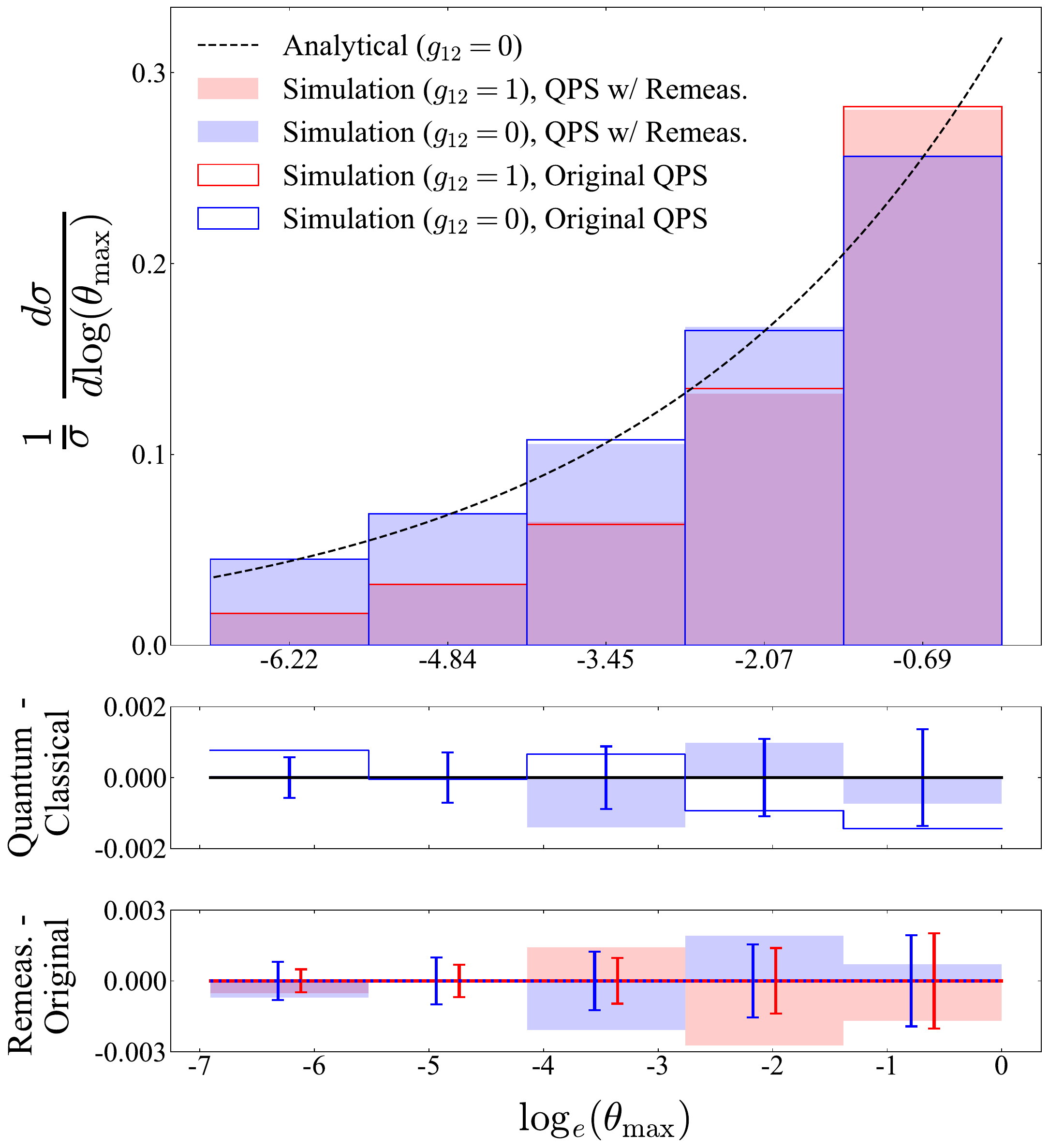}
    \end{subfigure}
    \captionsetup{font={normalsize}}
    \caption{Probability vs. $\log_e(\theta_{\text{max}})$ for 2, 3, 4, and 5 step simulations. As in \protect{\cref{fig:N_emissions}}, error bars represent $1\sigma$ ranges for the difference distributions. Simulation data (red, blue) is normalized such that probabilities are equal to bar height times bar width. For $g_{12}=0$ simulations (blue), the area of each bar is equal -- up to statistical deviations -- to the integral of the analytical curve (black) over the respective bin.}
    \label{fig:theta_max}
\end{figure*}

Figures~\ref{fig:N_emissions} and~\ref{fig:theta_max} present simulation results for $N=2, 3, 4, 5$ steps and compare the outputs between the original QPS and QPS with mid-circuit measurements\footnote{We have stopped at 5 steps due to the simulation time.  The present criteria for determining how many steps to use is that simulations with $10^5$ shots have to take fewer than 3 hours running naively without any parallelization on a 8 GB RAM Mac.  It would be possible to go a bit further with larger computing resources. For the remeasurement circuits, it took $\sim2.5$ hours to achieve $10^5$ shots for $g_{12}=0$ and $g_{12}=1$.  We note that classical conditioning is not fully implemented in \texttt{Qiskit} (it is not possible to do arbitrary classical calculations), so we have to apply an exponential number of classically-conditioned gates. }.
We have chosen two different observables for illustration.

First, \cref{fig:N_emissions} shows histograms of the total number of emissions $(E)$.
The main subplot illustrates the probability distributions of $E$ for classical MCMC (black), original QPS (filled bars), and QPS with mid-circuit measurements (solid edges), with both $g_{12}=0$ (blue) and $g_{12}=1$ (red).
The second subplot magnifies differences between the MCMC and $g_{12}=0$ simulation distributions, which are due to statistical noise and exhibit the expected deviations.
The third subplot magnifies differences between distributions obtained from original QPS and QPS with mid-circuit measurements, which are also within the expected statistical variations. 
With $10^5$ shots per simulation, typical statistical errors are on the order of $\sigma\sim\sqrt{\frac{\text{Pr}(E)}{10^5}}\lesssim \sqrt{\frac{1}{10^5}}\approx0.0032$.
Error bars shown in the second and third subplots of \cref{fig:N_emissions} are $1\sigma$ ranges for the difference distributions, and the simulation results exhibit deviations on the expected scale.
In other words, the second subplot shows that quantum simulations with coupling turned off $(g_{12}=0)$ agree with the classical MCMC algorithm, as expected.
Additionally, the two different versions of the quantum algorithm -- original and with mid-circuit measurements -- agree with one another.
Nevertheless, the classical $(g_{12}=0)$ and quantum $(g_{12}=1)$ algorithms yield fundamentally different results.

We briefly describe the qualitative features of \cref{fig:N_emissions}.
First, with the chosen parameters (\cref{eq:num_couplings,eq:params1,eq:splittings}), it turns out that
\begin{align}
    \text{Pr}(E=0, g_{12}=0) \,>\, \text{Pr}(E=0, g_{12}=1) \,.
\end{align}
Additionally, for $g_{12}=1$, the probability of $\phi$-emission is $1-\epsilon^{7/4\pi N}$, compared to $1-\epsilon^{5/4\pi N}$ for $g_{12}=0$.
Therefore, conditional on a $\phi$ particle being present in the system, $\phi$-emissions occur more frequently when $g_{12}=1$.
This explains why the probability of having just one emission for $g_{12}=1$ is so low compared to $g_{12}=0$. 
Finally, the exact shape of $E$ distributions depends on numerical parameter values, but the general shape exhibited in \cref{fig:N_emissions} --  increasing density with increasing $E$ up to a peak (which could be $E=N$), followed by a tail where density decreases as $E\rightarrow N$ -- is expected to hold for all $N$ and all parameter values.
The probability of emission at a given step is higher when there are more particles in the system, which explains why having just one or two emissions is less probable compared to having several emissions.
However at the tail end $(E\rightarrow N)$ of this trend, the distribution decreases slightly, because combinatorially there are more histories with $E=N-1$ then with $E=N$.

The second observable, \cref{fig:theta_max}, is the distribution of the ``hardest'' emission angle, which algorithmically corresponds to the first emission that occurred during the shower evolution.
The emission probability decreases exponentially with $\log(\theta)$, or linearly with opening angle $\theta$.
Algorithmically, this is because the  probability of first $f_i\rightarrow f_i\phi$ emission occuring at smaller angles (later steps) is just an exponential of the Sudakov factor (\cref{eq:sudakov_numerical}).
For $g_{12}=0$,
\begin{align}
 \begin{split}
    \text{Pr(First }\,&\text{emission at step}\,m) \\
    &\,=\, \left(1-\Delta_a(\theta_m)\right)\prod_{i=1}^{m-1}\Delta_a(\theta_m)\\
    &\,=\,\left(\Delta_a\right)^{m-1}\left(1-\Delta_a\right)  \,,
 \end{split}
\end{align}
and for $g_{12}=1$,
\begin{align}
 \begin{split}
    \text{Pr}&\text{(First emission at step}\,m) \\
    \,=\, &u\left(\Delta_a\right)^{m-1}\left(1-\Delta_a\right)\\
    + &(1-u)\left(\Delta_b\right)^{m-1}\left(1-\Delta_b\right) \,.
 \end{split}
\end{align}
\cref{fig:theta_max} shows the distribution of  $\log_e(\theta_{\text{max}})$, with probabilities displayed in the main subplot and differences displayed in the secondary subplots.
For this observable there is again a demonstrated difference in results between the classical $(g_{12}=0)$ and quantum $(g_{12}=1)$ algorithms.
Nevertheless, the third subplot illustrates that the two versions of QPS -- original and with mid-circuit measurements -- agree within expected statistical variations.

\section{Conclusions and Outlook}
\label{sec:conclusions}
In this paper, we have simplified the digital Quantum Parton Shower (QPS) algorithm presented in \cite{qps} by considering mid-circuit measurements and quantum gates that are dynamically selected based on these measurement results.
The QPS is an iterative ``quantum Markov Chain'' algorithm with $N$ steps, and by making a mid-circuit measurement on a subset of qubits at each step, subsequent multi-qubit controlled $R_y$ rotations in the original QPS can be replaced by dynamically selected single-qubit $R_y$ rotations.
In this case, the number of required $R_y$ rotations at each step is reduced by a factor of $N^2$.
Additionally, qubits measured mid-circuit can be reset to the initial $\ket{0}$ state and re-used during subsequent steps, which reduces qubit costs significantly.
The resulting \cref{alg:2} improves the quantum gate complexity from $\bO(N\cdot(N+n_I)^4\cdot\log(N+n_I)^2)$ to $\bO(N\cdot(N+n_I)^2\cdot\log(N+n_I)^2)$ (\cref{fig:gate_costs}), and the qubit complexity from $\bO(N\log(N+n_I))$ to $\bO(N+n_I)$ (\cref{fig:qubit_costs}), compared to the original algorithm (\cref{alg:1}).

We implement our quantum circuits using \texttt{Qiskit} (where dynamical quantum operations is a relatively new feature), and present results for $N=2,3,4,5$ steps.
We illustrate agreement between the original and improved versions of QPS, as well as agreement with classical MCMC simulations in the limit $g_{12}=0$, where QPS can be efficiently computed classically.
Errors are shown to be consistent with the expected statistical uncertainties in all cases.

More generally, we showed how adopting a hybrid quantum-classical computing platform can be used to make an originally quantum algorithm more efficient.
Recent studies \cite{dynamic_ibm, dynamic_raytheon} have demonstrated that dynamic/hybrid quantum computing is feasible, and even implemented shallow algorithms on current hardware.
As qubit design continues to improve, we expect to be able to execute more complicated hybrid algorithms such as QPS on real devices, and eventually be able to compute classically inaccessible physical observables.
Moreover, as dynamic quantum computing is an intrinsic component of quantum error correction, developing dynamic/hybrid computing platforms is likely necessary in order to realize fault-tolerant quantum computers. 
The improved QPS algorithm serves as an additional case for prioritizing development of dynamic computing platforms, as reduced qubit and gate complexities raise the potential for realizing QPS to compute classically inaccessible physical quantities much sooner.
It is likely that other digital quantum algorithms with similar features -- Markov Chain, or iterative algorithms where interferences exist within but not between different histories -- can benefit from employing a dynamic structure, and we encourage algorithm developers to consider this approach.

\section{Acknowledgements}
\label{sec:ack}
This work is supported by the U.S. Department of Energy, Office of Science under contract DE-AC02-05CH11231. In particular, support comes from Quantum Information Science Enabled Discovery (QuantISED) for High Energy Physics (KA2401032) and the Office of Advanced Scientific Computing Research (ASCR) through the Accelerated Research for Quantum Computing Program.   
\bibliographystyle{apsrev4-2}
\bibliography{main}

\begin{thebibliography}{33}%
\makeatletter
\providecommand \@ifxundefined [1]{%
 \@ifx{#1\undefined}
}%
\providecommand \@ifnum [1]{%
 \ifnum #1\expandafter \@firstoftwo
 \else \expandafter \@secondoftwo
 \fi
}%
\providecommand \@ifx [1]{%
 \ifx #1\expandafter \@firstoftwo
 \else \expandafter \@secondoftwo
 \fi
}%
\providecommand \natexlab [1]{#1}%
\providecommand \enquote  [1]{``#1''}%
\providecommand \bibnamefont  [1]{#1}%
\providecommand \bibfnamefont [1]{#1}%
\providecommand \citenamefont [1]{#1}%
\providecommand \href@noop [0]{\@secondoftwo}%
\providecommand \href [0]{\begingroup \@sanitize@url \@href}%
\providecommand \@href[1]{\@@startlink{#1}\@@href}%
\providecommand \@@href[1]{\endgroup#1\@@endlink}%
\providecommand \@sanitize@url [0]{\catcode `\\12\catcode `\$12\catcode
  `\&12\catcode `\#12\catcode `\^12\catcode `\_12\catcode `\%12\relax}%
\providecommand \@@startlink[1]{}%
\providecommand \@@endlink[0]{}%
\providecommand \url  [0]{\begingroup\@sanitize@url \@url }%
\providecommand \@url [1]{\endgroup\@href {#1}{\urlprefix }}%
\providecommand \urlprefix  [0]{URL }%
\providecommand \Eprint [0]{\href }%
\providecommand \doibase [0]{https://doi.org/}%
\providecommand \selectlanguage [0]{\@gobble}%
\providecommand \bibinfo  [0]{\@secondoftwo}%
\providecommand \bibfield  [0]{\@secondoftwo}%
\providecommand \translation [1]{[#1]}%
\providecommand \BibitemOpen [0]{}%
\providecommand \bibitemStop [0]{}%
\providecommand \bibitemNoStop [0]{.\EOS\space}%
\providecommand \EOS [0]{\spacefactor3000\relax}%
\providecommand \BibitemShut  [1]{\csname bibitem#1\endcsname}%
\let\auto@bib@innerbib\@empty
\bibitem [{\citenamefont {Nachman}\ \emph {et~al.}(2021)\citenamefont
  {Nachman}, \citenamefont {Provasoli}, \citenamefont {de~Jong},\ and\
  \citenamefont {Bauer}}]{qps}%
  \BibitemOpen
  \bibfield  {author} {\bibinfo {author} {\bibfnamefont {B.}~\bibnamefont
  {Nachman}}, \bibinfo {author} {\bibfnamefont {D.}~\bibnamefont {Provasoli}},
  \bibinfo {author} {\bibfnamefont {W.~A.}\ \bibnamefont {de~Jong}},\ and\
  \bibinfo {author} {\bibfnamefont {C.~W.}\ \bibnamefont {Bauer}},\ }\bibfield
  {journal} {\bibinfo  {journal} {Physical Review Letters}\ }\textbf {\bibinfo
  {volume} {126}},\ \href {https://doi.org/10.1103/physrevlett.126.062001}
  {10.1103/physrevlett.126.062001} (\bibinfo {year} {2021})\BibitemShut
  {NoStop}%
\bibitem [{\citenamefont {Jordan}\ \emph {et~al.}(2012)\citenamefont {Jordan},
  \citenamefont {Lee},\ and\ \citenamefont {Preskill}}]{Jordan:2012xnu}%
  \BibitemOpen
  \bibfield  {author} {\bibinfo {author} {\bibfnamefont {S.~P.}\ \bibnamefont
  {Jordan}}, \bibinfo {author} {\bibfnamefont {K.~S.~M.}\ \bibnamefont {Lee}},\
  and\ \bibinfo {author} {\bibfnamefont {J.}~\bibnamefont {Preskill}},\ }\href
  {https://doi.org/10.1126/science.1217069} {\bibfield  {journal} {\bibinfo
  {journal} {Science}\ }\textbf {\bibinfo {volume} {336}},\ \bibinfo {pages}
  {1130} (\bibinfo {year} {2012})},\ \Eprint {https://arxiv.org/abs/1111.3633}
  {arXiv:1111.3633 [quant-ph]} \BibitemShut {NoStop}%
\bibitem [{\citenamefont {Preskill}(2018{\natexlab{a}})}]{Preskill:2018fag}%
  \BibitemOpen
  \bibfield  {author} {\bibinfo {author} {\bibfnamefont {J.}~\bibnamefont
  {Preskill}},\ }\href {https://doi.org/10.22323/1.334.0024} {\bibfield
  {journal} {\bibinfo  {journal} {PoS}\ }\textbf {\bibinfo {volume}
  {LATTICE2018}},\ \bibinfo {pages} {024} (\bibinfo {year}
  {2018}{\natexlab{a}})},\ \Eprint {https://arxiv.org/abs/1811.10085}
  {arXiv:1811.10085 [hep-lat]} \BibitemShut {NoStop}%
\bibitem [{\citenamefont {Bauer}\ \emph {et~al.}(2021)\citenamefont {Bauer},
  \citenamefont {Freytsis},\ and\ \citenamefont {Nachman}}]{Bauer:2021gup}%
  \BibitemOpen
  \bibfield  {author} {\bibinfo {author} {\bibfnamefont {C.~W.}\ \bibnamefont
  {Bauer}}, \bibinfo {author} {\bibfnamefont {M.}~\bibnamefont {Freytsis}},\
  and\ \bibinfo {author} {\bibfnamefont {B.}~\bibnamefont {Nachman}},\ }\href
  {https://doi.org/10.1103/PhysRevLett.127.212001} {\bibfield  {journal}
  {\bibinfo  {journal} {Phys. Rev. Lett.}\ }\textbf {\bibinfo {volume} {127}},\
  \bibinfo {pages} {212001} (\bibinfo {year} {2021})},\ \Eprint
  {https://arxiv.org/abs/2102.05044} {arXiv:2102.05044 [hep-ph]} \BibitemShut
  {NoStop}%
\bibitem [{\citenamefont {Preskill}(2018{\natexlab{b}})}]{nisq}%
  \BibitemOpen
  \bibfield  {author} {\bibinfo {author} {\bibfnamefont {J.}~\bibnamefont
  {Preskill}},\ }\href {https://doi.org/10.22331/q-2018-08-06-79} {\bibfield
  {journal} {\bibinfo  {journal} {Quantum}\ }\textbf {\bibinfo {volume} {2}},\
  \bibinfo {pages} {79} (\bibinfo {year} {2018}{\natexlab{b}})}\BibitemShut
  {NoStop}%
\bibitem [{\citenamefont {ANIS}\ \emph {et~al.}(2021)\citenamefont {ANIS} \emph
  {et~al.}}]{Qiskit}%
  \BibitemOpen
  \bibfield  {author} {\bibinfo {author} {\bibfnamefont {M.~S.}\ \bibnamefont
  {ANIS}} \emph {et~al.},\ }\href {https://doi.org/10.5281/zenodo.2573505}
  {\bibinfo {title} {Qiskit: An open-source framework for quantum computing}}
  (\bibinfo {year} {2021})\BibitemShut {NoStop}%
\bibitem [{\citenamefont {Peruzzo}\ \emph {et~al.}(2014)\citenamefont
  {Peruzzo}, \citenamefont {McClean}, \citenamefont {Shadbolt}, \citenamefont
  {Yung}, \citenamefont {Zhou}, \citenamefont {Love}, \citenamefont
  {Aspuru-Guzik},\ and\ \citenamefont {O’Brien}}]{vqe}%
  \BibitemOpen
  \bibfield  {author} {\bibinfo {author} {\bibfnamefont {A.}~\bibnamefont
  {Peruzzo}}, \bibinfo {author} {\bibfnamefont {J.}~\bibnamefont {McClean}},
  \bibinfo {author} {\bibfnamefont {P.}~\bibnamefont {Shadbolt}}, \bibinfo
  {author} {\bibfnamefont {M.-H.}\ \bibnamefont {Yung}}, \bibinfo {author}
  {\bibfnamefont {X.-Q.}\ \bibnamefont {Zhou}}, \bibinfo {author}
  {\bibfnamefont {P.~J.}\ \bibnamefont {Love}}, \bibinfo {author}
  {\bibfnamefont {A.}~\bibnamefont {Aspuru-Guzik}},\ and\ \bibinfo {author}
  {\bibfnamefont {J.~L.}\ \bibnamefont {O’Brien}},\ }\bibfield  {journal}
  {\bibinfo  {journal} {Nature Communications}\ }\textbf {\bibinfo {volume}
  {5}},\ \href {https://doi.org/10.1038/ncomms5213} {10.1038/ncomms5213}
  (\bibinfo {year} {2014})\BibitemShut {NoStop}%
\bibitem [{\citenamefont {Nielsen}\ and\ \citenamefont
  {Chuang}(2010)}]{nielsen_chuang_2010}%
  \BibitemOpen
  \bibfield  {author} {\bibinfo {author} {\bibfnamefont {M.~A.}\ \bibnamefont
  {Nielsen}}\ and\ \bibinfo {author} {\bibfnamefont {I.~L.}\ \bibnamefont
  {Chuang}},\ }\href {https://doi.org/10.1017/CBO9780511976667} {\emph
  {\bibinfo {title} {Quantum Computation and Quantum Information: 10th
  Anniversary Edition}}}\ (\bibinfo  {publisher} {Cambridge University Press},\
  \bibinfo {year} {2010})\BibitemShut {NoStop}%
\bibitem [{\citenamefont {Ristè}\ \emph {et~al.}(2012)\citenamefont {Ristè},
  \citenamefont {Bultink}, \citenamefont {Lehnert},\ and\ \citenamefont
  {DiCarlo}}]{Riste2012}%
  \BibitemOpen
  \bibfield  {author} {\bibinfo {author} {\bibfnamefont {D.}~\bibnamefont
  {Ristè}}, \bibinfo {author} {\bibfnamefont {C.~C.}\ \bibnamefont {Bultink}},
  \bibinfo {author} {\bibfnamefont {K.~W.}\ \bibnamefont {Lehnert}},\ and\
  \bibinfo {author} {\bibfnamefont {L.}~\bibnamefont {DiCarlo}},\ }\bibfield
  {journal} {\bibinfo  {journal} {Physical Review Letters}\ }\textbf {\bibinfo
  {volume} {109}},\ \href {https://doi.org/10.1103/physrevlett.109.240502}
  {10.1103/physrevlett.109.240502} (\bibinfo {year} {2012})\BibitemShut
  {NoStop}%
\bibitem [{\citenamefont {Salath\'e}\ \emph {et~al.}(2018)\citenamefont
  {Salath\'e}, \citenamefont {Kurpiers}, \citenamefont {Karg}, \citenamefont
  {Lang}, \citenamefont {Andersen}, \citenamefont {Akin}, \citenamefont
  {Krinner}, \citenamefont {Eichler},\ and\ \citenamefont
  {Wallraff}}]{Salathe2018}%
  \BibitemOpen
  \bibfield  {author} {\bibinfo {author} {\bibfnamefont {Y.}~\bibnamefont
  {Salath\'e}}, \bibinfo {author} {\bibfnamefont {P.}~\bibnamefont {Kurpiers}},
  \bibinfo {author} {\bibfnamefont {T.}~\bibnamefont {Karg}}, \bibinfo {author}
  {\bibfnamefont {C.}~\bibnamefont {Lang}}, \bibinfo {author} {\bibfnamefont
  {C.~K.}\ \bibnamefont {Andersen}}, \bibinfo {author} {\bibfnamefont
  {A.}~\bibnamefont {Akin}}, \bibinfo {author} {\bibfnamefont {S.}~\bibnamefont
  {Krinner}}, \bibinfo {author} {\bibfnamefont {C.}~\bibnamefont {Eichler}},\
  and\ \bibinfo {author} {\bibfnamefont {A.}~\bibnamefont {Wallraff}},\ }\href
  {https://doi.org/10.1103/PhysRevApplied.9.034011} {\bibfield  {journal}
  {\bibinfo  {journal} {Phys. Rev. Applied}\ }\textbf {\bibinfo {volume} {9}},\
  \bibinfo {pages} {034011} (\bibinfo {year} {2018})}\BibitemShut {NoStop}%
\bibitem [{\citenamefont {Xiang}\ \emph {et~al.}(2020)\citenamefont {Xiang},
  \citenamefont {Zong}, \citenamefont {Sun}, \citenamefont {Zhan},
  \citenamefont {Fei}, \citenamefont {Dong}, \citenamefont {Run}, \citenamefont
  {Jia}, \citenamefont {Duan}, \citenamefont {Wu}, \citenamefont {Yin},\ and\
  \citenamefont {Guo}}]{Xiang2020}%
  \BibitemOpen
  \bibfield  {author} {\bibinfo {author} {\bibfnamefont {L.}~\bibnamefont
  {Xiang}}, \bibinfo {author} {\bibfnamefont {Z.}~\bibnamefont {Zong}},
  \bibinfo {author} {\bibfnamefont {Z.}~\bibnamefont {Sun}}, \bibinfo {author}
  {\bibfnamefont {Z.}~\bibnamefont {Zhan}}, \bibinfo {author} {\bibfnamefont
  {Y.}~\bibnamefont {Fei}}, \bibinfo {author} {\bibfnamefont {Z.}~\bibnamefont
  {Dong}}, \bibinfo {author} {\bibfnamefont {C.}~\bibnamefont {Run}}, \bibinfo
  {author} {\bibfnamefont {Z.}~\bibnamefont {Jia}}, \bibinfo {author}
  {\bibfnamefont {P.}~\bibnamefont {Duan}}, \bibinfo {author} {\bibfnamefont
  {J.}~\bibnamefont {Wu}}, \bibinfo {author} {\bibfnamefont {Y.}~\bibnamefont
  {Yin}},\ and\ \bibinfo {author} {\bibfnamefont {G.}~\bibnamefont {Guo}},\
  }\href {https://doi.org/10.1103/PhysRevApplied.14.014099} {\bibfield
  {journal} {\bibinfo  {journal} {Phys. Rev. Applied}\ }\textbf {\bibinfo
  {volume} {14}},\ \bibinfo {pages} {014099} (\bibinfo {year}
  {2020})}\BibitemShut {NoStop}%
\bibitem [{\citenamefont {Steffen}\ \emph {et~al.}(2013)\citenamefont
  {Steffen}, \citenamefont {Salathe}, \citenamefont {Oppliger}, \citenamefont
  {Kurpiers}, \citenamefont {Baur}, \citenamefont {Lang}, \citenamefont
  {Eichler}, \citenamefont {Puebla-Hellmann}, \citenamefont {Fedorov},\ and\
  \citenamefont {Wallraff}}]{Steffen2013}%
  \BibitemOpen
  \bibfield  {author} {\bibinfo {author} {\bibfnamefont {L.}~\bibnamefont
  {Steffen}}, \bibinfo {author} {\bibfnamefont {Y.}~\bibnamefont {Salathe}},
  \bibinfo {author} {\bibfnamefont {M.}~\bibnamefont {Oppliger}}, \bibinfo
  {author} {\bibfnamefont {P.}~\bibnamefont {Kurpiers}}, \bibinfo {author}
  {\bibfnamefont {M.}~\bibnamefont {Baur}}, \bibinfo {author} {\bibfnamefont
  {C.}~\bibnamefont {Lang}}, \bibinfo {author} {\bibfnamefont {C.}~\bibnamefont
  {Eichler}}, \bibinfo {author} {\bibfnamefont {G.}~\bibnamefont
  {Puebla-Hellmann}}, \bibinfo {author} {\bibfnamefont {A.}~\bibnamefont
  {Fedorov}},\ and\ \bibinfo {author} {\bibfnamefont {A.}~\bibnamefont
  {Wallraff}},\ }\href {https://doi.org/10.1038/nature12422} {\bibfield
  {journal} {\bibinfo  {journal} {Nature}\ }\textbf {\bibinfo {volume} {500}},\
  \bibinfo {pages} {319} (\bibinfo {year} {2013})}\BibitemShut {NoStop}%
\bibitem [{\citenamefont {Barrett}\ \emph {et~al.}(2004)\citenamefont
  {Barrett}, \citenamefont {Chiaverini}, \citenamefont {Schaetz}, \citenamefont
  {Britton}, \citenamefont {Itano}, \citenamefont {Jost}, \citenamefont
  {Knill}, \citenamefont {Langer}, \citenamefont {Leibfried}, \citenamefont
  {Ozeri},\ and\ \citenamefont {Wineland}}]{Barrett2004}%
  \BibitemOpen
  \bibfield  {author} {\bibinfo {author} {\bibfnamefont {M.~D.}\ \bibnamefont
  {Barrett}}, \bibinfo {author} {\bibfnamefont {J.}~\bibnamefont {Chiaverini}},
  \bibinfo {author} {\bibfnamefont {T.}~\bibnamefont {Schaetz}}, \bibinfo
  {author} {\bibfnamefont {J.}~\bibnamefont {Britton}}, \bibinfo {author}
  {\bibfnamefont {W.~M.}\ \bibnamefont {Itano}}, \bibinfo {author}
  {\bibfnamefont {J.~D.}\ \bibnamefont {Jost}}, \bibinfo {author}
  {\bibfnamefont {E.}~\bibnamefont {Knill}}, \bibinfo {author} {\bibfnamefont
  {C.}~\bibnamefont {Langer}}, \bibinfo {author} {\bibfnamefont
  {D.}~\bibnamefont {Leibfried}}, \bibinfo {author} {\bibfnamefont
  {R.}~\bibnamefont {Ozeri}},\ and\ \bibinfo {author} {\bibfnamefont {D.~J.}\
  \bibnamefont {Wineland}},\ }\href {https://doi.org/10.1038/nature02608}
  {\bibfield  {journal} {\bibinfo  {journal} {Nature}\ }\textbf {\bibinfo
  {volume} {429}},\ \bibinfo {pages} {737} (\bibinfo {year}
  {2004})}\BibitemShut {NoStop}%
\bibitem [{\citenamefont {Riebe}\ \emph {et~al.}(2004)\citenamefont {Riebe},
  \citenamefont {H{\"a}ffner}, \citenamefont {Roos}, \citenamefont
  {H{\"a}nsel}, \citenamefont {Benhelm}, \citenamefont {Lancaster},
  \citenamefont {K{\"o}rber}, \citenamefont {Becher}, \citenamefont
  {Schmidt-Kaler}, \citenamefont {James},\ and\ \citenamefont
  {Blatt}}]{Riebe2004}%
  \BibitemOpen
  \bibfield  {author} {\bibinfo {author} {\bibfnamefont {M.}~\bibnamefont
  {Riebe}}, \bibinfo {author} {\bibfnamefont {H.}~\bibnamefont {H{\"a}ffner}},
  \bibinfo {author} {\bibfnamefont {C.~F.}\ \bibnamefont {Roos}}, \bibinfo
  {author} {\bibfnamefont {W.}~\bibnamefont {H{\"a}nsel}}, \bibinfo {author}
  {\bibfnamefont {J.}~\bibnamefont {Benhelm}}, \bibinfo {author} {\bibfnamefont
  {G.~P.~T.}\ \bibnamefont {Lancaster}}, \bibinfo {author} {\bibfnamefont
  {T.~W.}\ \bibnamefont {K{\"o}rber}}, \bibinfo {author} {\bibfnamefont
  {C.}~\bibnamefont {Becher}}, \bibinfo {author} {\bibfnamefont
  {F.}~\bibnamefont {Schmidt-Kaler}}, \bibinfo {author} {\bibfnamefont
  {D.~F.~V.}\ \bibnamefont {James}},\ and\ \bibinfo {author} {\bibfnamefont
  {R.}~\bibnamefont {Blatt}},\ }\href {https://doi.org/10.1038/nature02570}
  {\bibfield  {journal} {\bibinfo  {journal} {Nature}\ }\textbf {\bibinfo
  {volume} {429}},\ \bibinfo {pages} {734} (\bibinfo {year}
  {2004})}\BibitemShut {NoStop}%
\bibitem [{\citenamefont {Chou}\ \emph {et~al.}(2018)\citenamefont {Chou},
  \citenamefont {Blumoff}, \citenamefont {Wang}, \citenamefont {Reinhold},
  \citenamefont {Axline}, \citenamefont {Gao}, \citenamefont {Frunzio},
  \citenamefont {Devoret}, \citenamefont {Jiang},\ and\ \citenamefont
  {Schoelkopf}}]{Chou2018}%
  \BibitemOpen
  \bibfield  {author} {\bibinfo {author} {\bibfnamefont {K.~S.}\ \bibnamefont
  {Chou}}, \bibinfo {author} {\bibfnamefont {J.~Z.}\ \bibnamefont {Blumoff}},
  \bibinfo {author} {\bibfnamefont {C.~S.}\ \bibnamefont {Wang}}, \bibinfo
  {author} {\bibfnamefont {P.~C.}\ \bibnamefont {Reinhold}}, \bibinfo {author}
  {\bibfnamefont {C.~J.}\ \bibnamefont {Axline}}, \bibinfo {author}
  {\bibfnamefont {Y.~Y.}\ \bibnamefont {Gao}}, \bibinfo {author} {\bibfnamefont
  {L.}~\bibnamefont {Frunzio}}, \bibinfo {author} {\bibfnamefont {M.~H.}\
  \bibnamefont {Devoret}}, \bibinfo {author} {\bibfnamefont {L.}~\bibnamefont
  {Jiang}},\ and\ \bibinfo {author} {\bibfnamefont {R.~J.}\ \bibnamefont
  {Schoelkopf}},\ }\href {https://doi.org/10.1038/s41586-018-0470-y} {\bibfield
   {journal} {\bibinfo  {journal} {Nature}\ }\textbf {\bibinfo {volume}
  {561}},\ \bibinfo {pages} {368} (\bibinfo {year} {2018})}\BibitemShut
  {NoStop}%
\bibitem [{\citenamefont {Furusawa}\ \emph {et~al.}(1998)\citenamefont
  {Furusawa}, \citenamefont {Sørensen}, \citenamefont {Braunstein},
  \citenamefont {Fuchs}, \citenamefont {Kimble},\ and\ \citenamefont
  {Polzik}}]{Furusawa1998}%
  \BibitemOpen
  \bibfield  {author} {\bibinfo {author} {\bibfnamefont {A.}~\bibnamefont
  {Furusawa}}, \bibinfo {author} {\bibfnamefont {J.~L.}\ \bibnamefont
  {Sørensen}}, \bibinfo {author} {\bibfnamefont {S.~L.}\ \bibnamefont
  {Braunstein}}, \bibinfo {author} {\bibfnamefont {C.~A.}\ \bibnamefont
  {Fuchs}}, \bibinfo {author} {\bibfnamefont {H.~J.}\ \bibnamefont {Kimble}},\
  and\ \bibinfo {author} {\bibfnamefont {E.~S.}\ \bibnamefont {Polzik}},\
  }\href {https://doi.org/10.1126/science.282.5389.706} {\bibfield  {journal}
  {\bibinfo  {journal} {Science}\ }\textbf {\bibinfo {volume} {282}},\ \bibinfo
  {pages} {706} (\bibinfo {year} {1998})},\ \Eprint
  {https://arxiv.org/abs/https://www.science.org/doi/pdf/10.1126/science.282.5389.706}
  {https://www.science.org/doi/pdf/10.1126/science.282.5389.706} \BibitemShut
  {NoStop}%
\bibitem [{\citenamefont {Rist{\`e}}\ \emph {et~al.}(2020)\citenamefont
  {Rist{\`e}}, \citenamefont {Govia}, \citenamefont {Donovan}, \citenamefont
  {Fallek}, \citenamefont {Kalfus}, \citenamefont {Brink}, \citenamefont
  {Bronn},\ and\ \citenamefont {Ohki}}]{Riste2020}%
  \BibitemOpen
  \bibfield  {author} {\bibinfo {author} {\bibfnamefont {D.}~\bibnamefont
  {Rist{\`e}}}, \bibinfo {author} {\bibfnamefont {L.~C.~G.}\ \bibnamefont
  {Govia}}, \bibinfo {author} {\bibfnamefont {B.}~\bibnamefont {Donovan}},
  \bibinfo {author} {\bibfnamefont {S.~D.}\ \bibnamefont {Fallek}}, \bibinfo
  {author} {\bibfnamefont {W.~D.}\ \bibnamefont {Kalfus}}, \bibinfo {author}
  {\bibfnamefont {M.}~\bibnamefont {Brink}}, \bibinfo {author} {\bibfnamefont
  {N.~T.}\ \bibnamefont {Bronn}},\ and\ \bibinfo {author} {\bibfnamefont
  {T.~A.}\ \bibnamefont {Ohki}},\ }\href
  {https://doi.org/10.1038/s41534-020-00304-y} {\bibfield  {journal} {\bibinfo
  {journal} {npj Quantum Information}\ }\textbf {\bibinfo {volume} {6}},\
  \bibinfo {pages} {71} (\bibinfo {year} {2020})}\BibitemShut {NoStop}%
\bibitem [{\citenamefont {Ofek}\ \emph {et~al.}(2016)\citenamefont {Ofek},
  \citenamefont {Petrenko}, \citenamefont {Heeres}, \citenamefont {Reinhold},
  \citenamefont {Leghtas}, \citenamefont {Vlastakis}, \citenamefont {Liu},
  \citenamefont {Frunzio}, \citenamefont {Girvin}, \citenamefont {Jiang},
  \citenamefont {Mirrahimi}, \citenamefont {Devoret},\ and\ \citenamefont
  {Schoelkopf}}]{Ofek2016}%
  \BibitemOpen
  \bibfield  {author} {\bibinfo {author} {\bibfnamefont {N.}~\bibnamefont
  {Ofek}}, \bibinfo {author} {\bibfnamefont {A.}~\bibnamefont {Petrenko}},
  \bibinfo {author} {\bibfnamefont {R.}~\bibnamefont {Heeres}}, \bibinfo
  {author} {\bibfnamefont {P.}~\bibnamefont {Reinhold}}, \bibinfo {author}
  {\bibfnamefont {Z.}~\bibnamefont {Leghtas}}, \bibinfo {author} {\bibfnamefont
  {B.}~\bibnamefont {Vlastakis}}, \bibinfo {author} {\bibfnamefont
  {Y.}~\bibnamefont {Liu}}, \bibinfo {author} {\bibfnamefont {L.}~\bibnamefont
  {Frunzio}}, \bibinfo {author} {\bibfnamefont {S.~M.}\ \bibnamefont {Girvin}},
  \bibinfo {author} {\bibfnamefont {L.}~\bibnamefont {Jiang}}, \bibinfo
  {author} {\bibfnamefont {M.}~\bibnamefont {Mirrahimi}}, \bibinfo {author}
  {\bibfnamefont {M.~H.}\ \bibnamefont {Devoret}},\ and\ \bibinfo {author}
  {\bibfnamefont {R.~J.}\ \bibnamefont {Schoelkopf}},\ }\href
  {https://doi.org/10.1038/nature18949} {\bibfield  {journal} {\bibinfo
  {journal} {Nature}\ }\textbf {\bibinfo {volume} {536}},\ \bibinfo {pages}
  {441} (\bibinfo {year} {2016})}\BibitemShut {NoStop}%
\bibitem [{\citenamefont {Campagne-Ibarcq}\ \emph {et~al.}(2013)\citenamefont
  {Campagne-Ibarcq}, \citenamefont {Flurin}, \citenamefont {Roch},
  \citenamefont {Darson}, \citenamefont {Morfin}, \citenamefont {Mirrahimi},
  \citenamefont {Devoret}, \citenamefont {Mallet},\ and\ \citenamefont
  {Huard}}]{CampagneIbarcq2013}%
  \BibitemOpen
  \bibfield  {author} {\bibinfo {author} {\bibfnamefont {P.}~\bibnamefont
  {Campagne-Ibarcq}}, \bibinfo {author} {\bibfnamefont {E.}~\bibnamefont
  {Flurin}}, \bibinfo {author} {\bibfnamefont {N.}~\bibnamefont {Roch}},
  \bibinfo {author} {\bibfnamefont {D.}~\bibnamefont {Darson}}, \bibinfo
  {author} {\bibfnamefont {P.}~\bibnamefont {Morfin}}, \bibinfo {author}
  {\bibfnamefont {M.}~\bibnamefont {Mirrahimi}}, \bibinfo {author}
  {\bibfnamefont {M.~H.}\ \bibnamefont {Devoret}}, \bibinfo {author}
  {\bibfnamefont {F.}~\bibnamefont {Mallet}},\ and\ \bibinfo {author}
  {\bibfnamefont {B.}~\bibnamefont {Huard}},\ }\href
  {https://doi.org/10.1103/PhysRevX.3.021008} {\bibfield  {journal} {\bibinfo
  {journal} {Phys. Rev. X}\ }\textbf {\bibinfo {volume} {3}},\ \bibinfo {pages}
  {021008} (\bibinfo {year} {2013})}\BibitemShut {NoStop}%
\bibitem [{\citenamefont {Minev}\ \emph {et~al.}(2019)\citenamefont {Minev},
  \citenamefont {Mundhada}, \citenamefont {Shankar}, \citenamefont {Reinhold},
  \citenamefont {Guti{\'e}rrez-J{\'a}uregui}, \citenamefont {Schoelkopf},
  \citenamefont {Mirrahimi}, \citenamefont {Carmichael},\ and\ \citenamefont
  {Devoret}}]{Minev2019}%
  \BibitemOpen
  \bibfield  {author} {\bibinfo {author} {\bibfnamefont {Z.~K.}\ \bibnamefont
  {Minev}}, \bibinfo {author} {\bibfnamefont {S.~O.}\ \bibnamefont {Mundhada}},
  \bibinfo {author} {\bibfnamefont {S.}~\bibnamefont {Shankar}}, \bibinfo
  {author} {\bibfnamefont {P.}~\bibnamefont {Reinhold}}, \bibinfo {author}
  {\bibfnamefont {R.}~\bibnamefont {Guti{\'e}rrez-J{\'a}uregui}}, \bibinfo
  {author} {\bibfnamefont {R.~J.}\ \bibnamefont {Schoelkopf}}, \bibinfo
  {author} {\bibfnamefont {M.}~\bibnamefont {Mirrahimi}}, \bibinfo {author}
  {\bibfnamefont {H.~J.}\ \bibnamefont {Carmichael}},\ and\ \bibinfo {author}
  {\bibfnamefont {M.~H.}\ \bibnamefont {Devoret}},\ }\href
  {https://doi.org/10.1038/s41586-019-1287-z} {\bibfield  {journal} {\bibinfo
  {journal} {Nature}\ }\textbf {\bibinfo {volume} {570}},\ \bibinfo {pages}
  {200} (\bibinfo {year} {2019})}\BibitemShut {NoStop}%
\bibitem [{\citenamefont {Pittman}\ \emph {et~al.}(2005)\citenamefont
  {Pittman}, \citenamefont {Jacobs},\ and\ \citenamefont
  {Franson}}]{Pittman2005}%
  \BibitemOpen
  \bibfield  {author} {\bibinfo {author} {\bibfnamefont {T.~B.}\ \bibnamefont
  {Pittman}}, \bibinfo {author} {\bibfnamefont {B.~C.}\ \bibnamefont
  {Jacobs}},\ and\ \bibinfo {author} {\bibfnamefont {J.~D.}\ \bibnamefont
  {Franson}},\ }\href {https://doi.org/10.1103/PhysRevA.71.052332} {\bibfield
  {journal} {\bibinfo  {journal} {Phys. Rev. A}\ }\textbf {\bibinfo {volume}
  {71}},\ \bibinfo {pages} {052332} (\bibinfo {year} {2005})}\BibitemShut
  {NoStop}%
\bibitem [{\citenamefont {Chiaverini}\ \emph {et~al.}(2004)\citenamefont
  {Chiaverini}, \citenamefont {Leibfried}, \citenamefont {Schaetz},
  \citenamefont {Barrett}, \citenamefont {Blakestad}, \citenamefont {Britton},
  \citenamefont {Itano}, \citenamefont {Jost}, \citenamefont {Knill},
  \citenamefont {Langer}, \citenamefont {Ozeri},\ and\ \citenamefont
  {Wineland}}]{Chiaverini2004}%
  \BibitemOpen
  \bibfield  {author} {\bibinfo {author} {\bibfnamefont {J.}~\bibnamefont
  {Chiaverini}}, \bibinfo {author} {\bibfnamefont {D.}~\bibnamefont
  {Leibfried}}, \bibinfo {author} {\bibfnamefont {T.}~\bibnamefont {Schaetz}},
  \bibinfo {author} {\bibfnamefont {M.~D.}\ \bibnamefont {Barrett}}, \bibinfo
  {author} {\bibfnamefont {R.~B.}\ \bibnamefont {Blakestad}}, \bibinfo {author}
  {\bibfnamefont {J.}~\bibnamefont {Britton}}, \bibinfo {author} {\bibfnamefont
  {W.~M.}\ \bibnamefont {Itano}}, \bibinfo {author} {\bibfnamefont {J.~D.}\
  \bibnamefont {Jost}}, \bibinfo {author} {\bibfnamefont {E.}~\bibnamefont
  {Knill}}, \bibinfo {author} {\bibfnamefont {C.}~\bibnamefont {Langer}},
  \bibinfo {author} {\bibfnamefont {R.}~\bibnamefont {Ozeri}},\ and\ \bibinfo
  {author} {\bibfnamefont {D.~J.}\ \bibnamefont {Wineland}},\ }\href
  {https://doi.org/10.1038/nature03074} {\bibfield  {journal} {\bibinfo
  {journal} {Nature}\ }\textbf {\bibinfo {volume} {432}},\ \bibinfo {pages}
  {602} (\bibinfo {year} {2004})}\BibitemShut {NoStop}%
\bibitem [{\citenamefont {Vijay}\ \emph {et~al.}(2012)\citenamefont {Vijay},
  \citenamefont {Macklin}, \citenamefont {Slichter}, \citenamefont {Weber},
  \citenamefont {Murch}, \citenamefont {Naik}, \citenamefont {Korotkov},\ and\
  \citenamefont {Siddiqi}}]{Vijay2012}%
  \BibitemOpen
  \bibfield  {author} {\bibinfo {author} {\bibfnamefont {R.}~\bibnamefont
  {Vijay}}, \bibinfo {author} {\bibfnamefont {C.}~\bibnamefont {Macklin}},
  \bibinfo {author} {\bibfnamefont {D.~H.}\ \bibnamefont {Slichter}}, \bibinfo
  {author} {\bibfnamefont {S.~J.}\ \bibnamefont {Weber}}, \bibinfo {author}
  {\bibfnamefont {K.~W.}\ \bibnamefont {Murch}}, \bibinfo {author}
  {\bibfnamefont {R.}~\bibnamefont {Naik}}, \bibinfo {author} {\bibfnamefont
  {A.~N.}\ \bibnamefont {Korotkov}},\ and\ \bibinfo {author} {\bibfnamefont
  {I.}~\bibnamefont {Siddiqi}},\ }\href {https://doi.org/10.1038/nature11505}
  {\bibfield  {journal} {\bibinfo  {journal} {Nature}\ }\textbf {\bibinfo
  {volume} {490}},\ \bibinfo {pages} {77} (\bibinfo {year} {2012})}\BibitemShut
  {NoStop}%
\bibitem [{\citenamefont {Hu}\ \emph {et~al.}(2019)\citenamefont {Hu},
  \citenamefont {Ma}, \citenamefont {Cai}, \citenamefont {Mu}, \citenamefont
  {Xu}, \citenamefont {Wang}, \citenamefont {Wu}, \citenamefont {Wang},
  \citenamefont {Song}, \citenamefont {Zou}, \citenamefont {Girvin},
  \citenamefont {Duan},\ and\ \citenamefont {Sun}}]{Hu2019}%
  \BibitemOpen
  \bibfield  {author} {\bibinfo {author} {\bibfnamefont {L.}~\bibnamefont
  {Hu}}, \bibinfo {author} {\bibfnamefont {Y.}~\bibnamefont {Ma}}, \bibinfo
  {author} {\bibfnamefont {W.}~\bibnamefont {Cai}}, \bibinfo {author}
  {\bibfnamefont {X.}~\bibnamefont {Mu}}, \bibinfo {author} {\bibfnamefont
  {Y.}~\bibnamefont {Xu}}, \bibinfo {author} {\bibfnamefont {W.}~\bibnamefont
  {Wang}}, \bibinfo {author} {\bibfnamefont {Y.}~\bibnamefont {Wu}}, \bibinfo
  {author} {\bibfnamefont {H.}~\bibnamefont {Wang}}, \bibinfo {author}
  {\bibfnamefont {Y.~P.}\ \bibnamefont {Song}}, \bibinfo {author}
  {\bibfnamefont {C.-L.}\ \bibnamefont {Zou}}, \bibinfo {author} {\bibfnamefont
  {S.~M.}\ \bibnamefont {Girvin}}, \bibinfo {author} {\bibfnamefont {L.-M.}\
  \bibnamefont {Duan}},\ and\ \bibinfo {author} {\bibfnamefont
  {L.}~\bibnamefont {Sun}},\ }\href {https://doi.org/10.1038/s41567-018-0414-3}
  {\bibfield  {journal} {\bibinfo  {journal} {Nature Physics}\ }\textbf
  {\bibinfo {volume} {15}},\ \bibinfo {pages} {503} (\bibinfo {year}
  {2019})}\BibitemShut {NoStop}%
\bibitem [{\citenamefont {Andersen}\ \emph {et~al.}(2019)\citenamefont
  {Andersen}, \citenamefont {Remm}, \citenamefont {Lazar}, \citenamefont
  {Krinner}, \citenamefont {Heinsoo}, \citenamefont {Besse}, \citenamefont
  {Gabureac}, \citenamefont {Wallraff},\ and\ \citenamefont
  {Eichler}}]{Andersen2019}%
  \BibitemOpen
  \bibfield  {author} {\bibinfo {author} {\bibfnamefont {C.~K.}\ \bibnamefont
  {Andersen}}, \bibinfo {author} {\bibfnamefont {A.}~\bibnamefont {Remm}},
  \bibinfo {author} {\bibfnamefont {S.}~\bibnamefont {Lazar}}, \bibinfo
  {author} {\bibfnamefont {S.}~\bibnamefont {Krinner}}, \bibinfo {author}
  {\bibfnamefont {J.}~\bibnamefont {Heinsoo}}, \bibinfo {author} {\bibfnamefont
  {J.-C.}\ \bibnamefont {Besse}}, \bibinfo {author} {\bibfnamefont
  {M.}~\bibnamefont {Gabureac}}, \bibinfo {author} {\bibfnamefont
  {A.}~\bibnamefont {Wallraff}},\ and\ \bibinfo {author} {\bibfnamefont
  {C.}~\bibnamefont {Eichler}},\ }\href
  {https://doi.org/10.1038/s41534-019-0185-4} {\bibfield  {journal} {\bibinfo
  {journal} {npj Quantum Information}\ }\textbf {\bibinfo {volume} {5}},\
  \bibinfo {pages} {69} (\bibinfo {year} {2019})}\BibitemShut {NoStop}%
\bibitem [{\citenamefont {Bultink}\ \emph {et~al.}(2020)\citenamefont
  {Bultink}, \citenamefont {O’Brien}, \citenamefont {Vollmer}, \citenamefont
  {Muthusubramanian}, \citenamefont {Beekman}, \citenamefont {Rol},
  \citenamefont {Fu}, \citenamefont {Tarasinski}, \citenamefont {Ostroukh},
  \citenamefont {Varbanov}, \citenamefont {Bruno},\ and\ \citenamefont
  {DiCarlo}}]{Bultink2020}%
  \BibitemOpen
  \bibfield  {author} {\bibinfo {author} {\bibfnamefont {C.~C.}\ \bibnamefont
  {Bultink}}, \bibinfo {author} {\bibfnamefont {T.~E.}\ \bibnamefont
  {O’Brien}}, \bibinfo {author} {\bibfnamefont {R.}~\bibnamefont {Vollmer}},
  \bibinfo {author} {\bibfnamefont {N.}~\bibnamefont {Muthusubramanian}},
  \bibinfo {author} {\bibfnamefont {M.~W.}\ \bibnamefont {Beekman}}, \bibinfo
  {author} {\bibfnamefont {M.~A.}\ \bibnamefont {Rol}}, \bibinfo {author}
  {\bibfnamefont {X.}~\bibnamefont {Fu}}, \bibinfo {author} {\bibfnamefont
  {B.}~\bibnamefont {Tarasinski}}, \bibinfo {author} {\bibfnamefont
  {V.}~\bibnamefont {Ostroukh}}, \bibinfo {author} {\bibfnamefont
  {B.}~\bibnamefont {Varbanov}}, \bibinfo {author} {\bibfnamefont
  {A.}~\bibnamefont {Bruno}},\ and\ \bibinfo {author} {\bibfnamefont
  {L.}~\bibnamefont {DiCarlo}},\ }\href
  {https://doi.org/10.1126/sciadv.aay3050} {\bibfield  {journal} {\bibinfo
  {journal} {Science Advances}\ }\textbf {\bibinfo {volume} {6}},\ \bibinfo
  {pages} {eaay3050} (\bibinfo {year} {2020})},\ \Eprint
  {https://arxiv.org/abs/https://www.science.org/doi/pdf/10.1126/sciadv.aay3050}
  {https://www.science.org/doi/pdf/10.1126/sciadv.aay3050} \BibitemShut
  {NoStop}%
\bibitem [{\citenamefont {Córcoles}\ \emph {et~al.}(2021)\citenamefont
  {Córcoles}, \citenamefont {Takita}, \citenamefont {Inoue}, \citenamefont
  {Lekuch}, \citenamefont {Minev}, \citenamefont {Chow},\ and\ \citenamefont
  {Gambetta}}]{dynamic_ibm}%
  \BibitemOpen
  \bibfield  {author} {\bibinfo {author} {\bibfnamefont {A.~D.}\ \bibnamefont
  {Córcoles}}, \bibinfo {author} {\bibfnamefont {M.}~\bibnamefont {Takita}},
  \bibinfo {author} {\bibfnamefont {K.}~\bibnamefont {Inoue}}, \bibinfo
  {author} {\bibfnamefont {S.}~\bibnamefont {Lekuch}}, \bibinfo {author}
  {\bibfnamefont {Z.~K.}\ \bibnamefont {Minev}}, \bibinfo {author}
  {\bibfnamefont {J.~M.}\ \bibnamefont {Chow}},\ and\ \bibinfo {author}
  {\bibfnamefont {J.~M.}\ \bibnamefont {Gambetta}},\ }\bibfield  {journal}
  {\bibinfo  {journal} {Physical Review Letters}\ }\textbf {\bibinfo {volume}
  {127}},\ \href {https://doi.org/10.1103/physrevlett.127.100501}
  {10.1103/physrevlett.127.100501} (\bibinfo {year} {2021})\BibitemShut
  {NoStop}%
\bibitem [{\citenamefont {Bepari}\ \emph {et~al.}(2021)\citenamefont {Bepari},
  \citenamefont {Malik}, \citenamefont {Spannowsky},\ and\ \citenamefont
  {Williams}}]{Bepari_2021}%
  \BibitemOpen
  \bibfield  {author} {\bibinfo {author} {\bibfnamefont {K.}~\bibnamefont
  {Bepari}}, \bibinfo {author} {\bibfnamefont {S.}~\bibnamefont {Malik}},
  \bibinfo {author} {\bibfnamefont {M.}~\bibnamefont {Spannowsky}},\ and\
  \bibinfo {author} {\bibfnamefont {S.}~\bibnamefont {Williams}},\ }\bibfield
  {journal} {\bibinfo  {journal} {Physical Review D}\ }\textbf {\bibinfo
  {volume} {103}},\ \href {https://doi.org/10.1103/physrevd.103.076020}
  {10.1103/physrevd.103.076020} (\bibinfo {year} {2021})\BibitemShut {NoStop}%
\bibitem [{\citenamefont {Williams}\ \emph {et~al.}(2021)\citenamefont
  {Williams}, \citenamefont {Malik}, \citenamefont {Spannowsky},\ and\
  \citenamefont {Bepari}}]{qps_random_walk}%
  \BibitemOpen
  \bibfield  {author} {\bibinfo {author} {\bibfnamefont {S.}~\bibnamefont
  {Williams}}, \bibinfo {author} {\bibfnamefont {S.}~\bibnamefont {Malik}},
  \bibinfo {author} {\bibfnamefont {M.}~\bibnamefont {Spannowsky}},\ and\
  \bibinfo {author} {\bibfnamefont {K.}~\bibnamefont {Bepari}},\ }\href
  {https://doi.org/10.48550/ARXIV.2109.13975} {\bibinfo {title} {A quantum walk
  approach to simulating parton showers}} (\bibinfo {year} {2021})\BibitemShut
  {NoStop}%
\bibitem [{\citenamefont {Schollwöck}(2011)}]{mps1}%
  \BibitemOpen
  \bibfield  {author} {\bibinfo {author} {\bibfnamefont {U.}~\bibnamefont
  {Schollwöck}},\ }\href {https://doi.org/10.1016/j.aop.2010.09.012}
  {\bibfield  {journal} {\bibinfo  {journal} {Annals of Physics}\ }\textbf
  {\bibinfo {volume} {326}},\ \bibinfo {pages} {96–192} (\bibinfo {year}
  {2011})}\BibitemShut {NoStop}%
\bibitem [{\citenamefont {Vidal}(2003)}]{mps2}%
  \BibitemOpen
  \bibfield  {author} {\bibinfo {author} {\bibfnamefont {G.}~\bibnamefont
  {Vidal}},\ }\bibfield  {journal} {\bibinfo  {journal} {Physical Review
  Letters}\ }\textbf {\bibinfo {volume} {91}},\ \href
  {https://doi.org/10.1103/physrevlett.91.147902}
  {10.1103/physrevlett.91.147902} (\bibinfo {year} {2003})\BibitemShut
  {NoStop}%
\bibitem [{\citenamefont {Ryan}\ \emph {et~al.}(2017)\citenamefont {Ryan},
  \citenamefont {Johnson}, \citenamefont {Ristè}, \citenamefont {Donovan},\
  and\ \citenamefont {Ohki}}]{dynamic_raytheon}%
  \BibitemOpen
  \bibfield  {author} {\bibinfo {author} {\bibfnamefont {C.~A.}\ \bibnamefont
  {Ryan}}, \bibinfo {author} {\bibfnamefont {B.~R.}\ \bibnamefont {Johnson}},
  \bibinfo {author} {\bibfnamefont {D.}~\bibnamefont {Ristè}}, \bibinfo
  {author} {\bibfnamefont {B.}~\bibnamefont {Donovan}},\ and\ \bibinfo {author}
  {\bibfnamefont {T.~A.}\ \bibnamefont {Ohki}},\ }\href
  {https://doi.org/10.1063/1.5006525} {\bibfield  {journal} {\bibinfo
  {journal} {Review of Scientific Instruments}\ }\textbf {\bibinfo {volume}
  {88}},\ \bibinfo {pages} {104703} (\bibinfo {year} {2017})}\BibitemShut
  {NoStop}%
\bibitem [{\citenamefont {Cuccaro}\ \emph {et~al.}(2004)\citenamefont
  {Cuccaro}, \citenamefont {Draper}, \citenamefont {Kutin},\ and\ \citenamefont
  {Moulton}}]{cuccaro2004new}%
  \BibitemOpen
  \bibfield  {author} {\bibinfo {author} {\bibfnamefont {S.~A.}\ \bibnamefont
  {Cuccaro}}, \bibinfo {author} {\bibfnamefont {T.~G.}\ \bibnamefont {Draper}},
  \bibinfo {author} {\bibfnamefont {S.~A.}\ \bibnamefont {Kutin}},\ and\
  \bibinfo {author} {\bibfnamefont {D.~P.}\ \bibnamefont {Moulton}},\
  }\href@noop {} {\bibinfo {title} {A new quantum ripple-carry addition
  circuit}} (\bibinfo {year} {2004}),\ \Eprint
  {https://arxiv.org/abs/quant-ph/0410184} {arXiv:quant-ph/0410184 [quant-ph]}
  \BibitemShut {NoStop}%
\end{thebibliography}%
\clearpage

\appendix
\section{Details of the original Quantum Parton Shower (QPS) algorithm}
\label{app:1}
This appendix provides details on the original QPS algorithm presented in \cite{qps}.

\subsubsection{Particle state \texorpdfstring{$\ket{p}$}{p}}
\label{subsubsec:preg}
This register consists of $N+n_I$ $3$-qubit sub-registers, one for each initial particle, and one for each emission step,
\begin{align}
    \ket{p} \,=\, \ket{p_1}^{\otimes3} \otimes \dots \otimes \ket{p_{n_I+N}}^{\otimes3}\,.
    \label{eq:preg}
\end{align}
We use three qubits to encode each particle, as there are six different types of particles ($f_1$, $f_2$, $\bar{f}_1$, $\bar{f}_2$, $\phi$, and None) in our model.
The exact encoding is relevant, and we use the following,
\begin{align}
    \ket{p_i} \,=\, \left(\begin{matrix}000\\001\\010\\011\\100\\101\\110\\111\end{matrix}\right) \,=\, \left(\begin{matrix}\text{None}\\\phi\\\\\\f_1/f_a\\f_2/f_b\\\bar{f}_1/\bar{f}_a\\\bar{f}_2/\bar{f}_b\end{matrix}\right)\,.
    \label{eq:p_state}
\end{align}
Therefore, operations conditioned on whether a particle is a fermion are controlled by just the first qubit, and operations conditioned on whether a fermion is type-$a$ or type-$b$ are controlled by just the first and third qubits.
Note that two computational basis states are extraneous.

\subsubsection{Emission history \texorpdfstring{$\ket{h}$}{h}}
\label{subsubsec:hreg}
In original QPS, $\ket{h}$ encodes the location of emission at each step.
In particular, at step $m$, $\ket{h_m}$ stores a binary number between $0$ and $n_I+m$ that specifies which particle emitted at that step.
The $\ket{0}$ state encodes that no emission occured.
Therefore, each sub-register $\ket{h_m}$ consists of $\ceil{\log_2(n_I+m+1)}$ qubits, and in total $\ket{h}$ consists of 
\begin{align}
    \sum_{m=0}^{N-1}(\ceil{\log_2(n_I+m+1)})\sim\bO(N\log_2(n_I+N))
\end{align} 
qubits.

\subsubsection{Emission \texorpdfstring{$\ket{e}$}{e}}
\label{subsubsec:ereg}
The emission register $\ket{e}$ stores a boolean that specifies whether an emission occurred at a given step.
It is straightforward to uncompute $\ket{e}$ after each step, so just one qubit is sufficient to represent $\ket{e}$.

\subsubsection{Number registers \texorpdfstring{$\ket{n_{\phi}}, \ket{n_{a}}, \ket{n_{b}}$}{nphi, na, nb}}
\label{subsubsec:nreg}
The number registers are used to count the number of each particle type at each step.
In particular, at step $m$ each of $\ket{n_{\phi}}, \ket{n_{a}}, \ket{n_{b}}$ stores a binary number between $0$ and $n_I+m$, the maximum possible number of particles at step $m$.
Also note that the total is constrained by
\begin{align}
    0 \leq n_{\phi}+n_a+n_b \leq n_I+m \,.
\end{align}
All three number registers are uncomputed after each step and can be re-used for subsequent steps.
Therefore, for an $N$-step algorithm, each number register consists of $\ceil{\log_2(n_I+N)}$ qubits, for a total of $3\ceil{\log_2(n_I+N)}$ qubits between $\ket{n_{\phi}}, \ket{n_{a}}$, and $\ket{n_{b}}$.

Having set up the six quantum registers shown in \cref{fig:qps_circ_orig}, we now describe each gate in the circuit.

\subsubsection{\texorpdfstring{$R^{(m)}$}{Rm} basis rotation}
\label{subsubsec:Rgate}
As described in \cref{subsec:basis}, we rotate fermion states from the $1/2$ basis into the $a/b$ basis by applying unitary $U$ (\cref{eq:basis_rotation}).
Given our particle state representation \ref{eq:p_state}, rotating a single particle represented by three qubits entails applying the following unitary gate:
\begin{align}
    R \,=\, \left(\begin{matrix}I & 0 & 0 & 0\\
                                0 & I & 0 & 0\\
                                0 & 0 & U & 0\\
                                0 & 0 & 0 & U
                  \end{matrix}\right) \,,
    \label{eq:basis_rotation}
\end{align}
where $I$ and $U$ are $2\times2$ matrices.
Therefore, to rotate to complete particle register $\ket{p}$ at step $m$, apply the product gate
\begin{align}
    R^{\otimes(n_I+m)} \,.
    \label{eq:r_prodgate}
\end{align}
The gate in \cref{eq:basis_rotation} is just a controlled-$U$ gate, where $U$ is applied to the rightmost qubit controlled on the leftmost qubit, in the particle encoding (\cref{eq:p_state}).
Therefore, applying \cref{eq:r_prodgate} at the beginning and end of step $m$ involves applying $(n_I+m)$ controlled-$U$ gates, each of which can be decomposed into two CNOTs.

\subsubsection{\texorpdfstring{$U_{\text{count}}$}{Ucount} particle counting}
\label{subsubsec:Ucount_old}
\begin{figure}[htp]
 \begin{adjustbox}{width=1.0\columnwidth}
    \begin{tikzcd}[column sep=0.12cm, row sep=0.12cm]
    \lstick{\ket{p_1}} & \qwbundle \qw & \qw & \qw & \qw & \qw & \ \ldots\ & \measure{\mbox{$\phi$}} \vqw{3} & \measure{\mbox{$f_a$}} \vqw{4} & \measure{\mbox{$f_b$}} \vqw{5} & \qw & \\
    \vdots &&&&&&\iddots&&&\\
    \lstick{\ket{p_{n_I+m}}} & \qwbundle \qw &  \measure{\mbox{$\phi$}} \vqw{1} & \measure{\mbox{$f_a$}} \vqw{2} & \measure{\mbox{$f_b$}} \vqw{3} & \qw & \ \ldots\ \qw & \qw & \qw & \qw & \qw \\
    \lstick{\ket{n_{\phi}}} & \qwbundle \qw & \gate{+} & \qw & \qw & \qw & \ \ldots\ \qw & \gate{+} & \qw & \qw & \qw\\
    \lstick{\ket{n_a}} & \qwbundle \qw  & \qw & \gate{+} & \qw & \qw & \ \ldots\ \qw & \qw & \gate{+} & \qw & \qw \\ 
    \lstick{\ket{n_b}} & \qwbundle \qw  & \qw & \qw & \gate{+} & \qw & \ \ldots\ \qw & \qw & \qw & \gate{+} & \qw
    \end{tikzcd}
 \end{adjustbox}
 \caption{Complete illustration of the $U_{\text{count}}$ gate.}
 \label{fig:Ucount_old}
\end{figure}
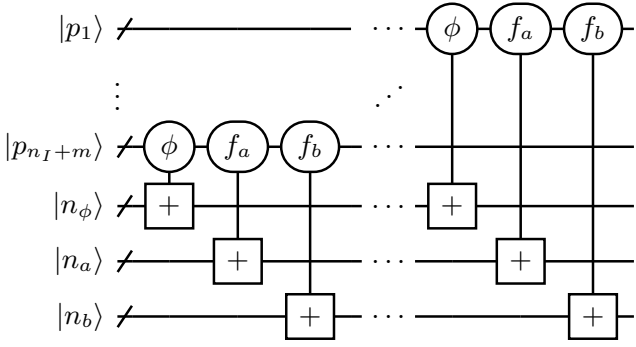

\begin{figure}[htp]
  \begin{adjustbox}{width=1.0\columnwidth}
    \begin{tikzcd}[column sep=0.24cm, row sep=0.52cm]
        \lstick{$\ket{p_i}^{(0)}$} & \ctrl{2} & \qw & \ctrl{2} & \qw & \rstick{$\ket{p_i}^{(0)}$}\\
        \lstick{$\ket{p_i}^{(1)}$} & \qw & \qw & \qw & \qw & \rstick{$\ket{p_i}^{(1)}$}\\
        \lstick{$\ket{p_i}^{(2)}$} & \octrl{1} & \qw & \octrl{1} & \qw & \rstick{$\ket{p_i}^{(2)}$}\\
        \lstick{\ket{0}} & \targ{} & \gate[3, disable auto height]{\rotatebox{0}{\textsc{Ripple Adder}}} \gateinput[1]{$c$} & \targ{} & \qw & \rstick{\ket{0}} \\
        \lstick{\ket{0}} & \qwbundle{\ell} & & \qw & \qw & \rstick{\ket{0}} \\
        \lstick{\ket{n_{\phi}}} & \qwbundle{\ell} & \gateoutput[1]{$\blacktriangleleft$} & \qw & \qw & \rstick{$\ket{n_{\phi} + c}$} \\
    \end{tikzcd}
  \end{adjustbox}
 \caption{Increment gate controlled on $\ket{p=\phi}$. First, a multi-control gate that encodes whether $\ket{p_i}$ is a $\phi$ onto an ancilla $\ket{c}$ is applied. Then, a Ripple Adder \cite{cuccaro2004new} with $\ket{c}$ as an input carry, and an additional ancillary register initialized to $\ket{0}$ is applied. Therefore, $\ket{n_{\phi}}$ is mapped to $\ket{n_{\phi}+1}$ if $c=1$, i.e. if $\ket{p_i}$ is a $\phi$. Increment gates controlled on $\ket{p=f_a}$ $\ket{p=f_b}$ are identical other than the control setting (see \protect{\ref{eq:p_state}}).}
 \label{fig:increment}
\end{figure}
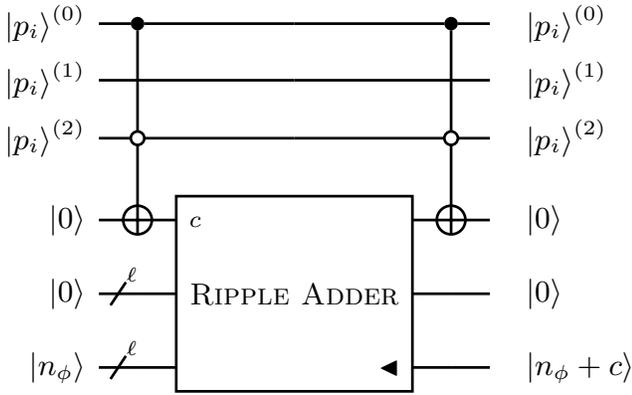

The counting gate maps the particle state $\ket{p}$ at step $m$ to the number of each particle, which is stored in $\ket{n_{\phi}}, \ket{n_{a}}, \ket{n_{b}}$.
Note that we count fermions and anti-fermions of the same type together, as this distinction does not affect emission probabilities.
For each particle $\ket{p_i}$ in the particle state $\ket{p}$, we apply an increment gate $\boxed{+}$ controlled on particle type to each of $\ket{n_{\phi}}, \ket{n_{a}}, \ket{n_{b}}$, as illustrated in \cref{fig:Ucount_old}
The increment gate controlled on $\phi$, $f_a$, and $f_b$ implements respective transformations
\begin{align}
    \ket{p_i}\ket{n_{\phi}} &\rightarrow \ket{p_i=\phi}\ket{n_{\phi}+1} + \ket{p_i\neq\phi}\ket{n_{\phi}}\\
    \ket{p_i}\ket{n_{a}} &\rightarrow \ket{p_i=f_a}\ket{n_{a}+1} + \ket{p_i\neq f_a}\ket{n_{a}}\\
    \ket{p_i}\ket{n_{b}} &\rightarrow \ket{p_i=f_b}\ket{n_{b}+1} + \ket{p_i\neq f_b}\ket{n_{b}}
    \,,
\end{align}
and is illustrated in \cref{fig:increment}.
In total, $U_{\text{count}}$ consists of $3(n_I+m)$ controlled-incrementers, each of which requires $\bO(\ceil{\log_2(n_I+m+1)})$ gates and $\ceil{\log_2(n_I+m+1)}$ re-usable ancillas.
Thus, the total gate complexity of $U_{\text{count}}$ is $\bO((n_I+m)\cdot\ceil{\log_2(n_I+m+1)})$.

\subsubsection{\texorpdfstring{$U_{e}$}{Ue} emission update}
\label{subsubsec:Ue_old}
At each step, the total probability that an emission occurs depends on the scale $\theta_m$, splitting functions $P_f(\theta_m), P_{\phi}(\theta_m)$, and numbers of each type of particle $\ket{n_a}, \ket{n_b}, \ket{n_{\phi}}$.
Let
\begin{align}
    \Delta_a(\theta_m), \,\Delta_b(\theta_m), \,\Delta_{\phi}(\theta_m)
\end{align}
denote the probability of \textit{no emission} from a particular $f_a$, $f_b$, or $\phi$ particle, respectively. 
(These are the Sudakov factors defined by \cref{eq:sudakov}.)
Then, the total probability of having \textit{no emission} at step $m$ is
\begin{align}
    &\text{Pr(No }\text{emission)} \nonumber\\
    &\,\equiv\, \Delta^{(m)} \,=\, \Delta_a(\theta_m)^{n_a}\Delta_b(\theta_m)^{n_b}\Delta_{\phi}(\theta_m)^{n_{\phi}}\,,
\end{align}
so that the probability of emission is
\begin{align}
    \text{Pr(Emit at step }m) \,=\, 1-\Delta^{(m)}
    \,.
\end{align}
Note that $\ket{n_a}, \ket{n_b}, \ket{n_{\phi}}$ are quantum superpositions while $\theta_m, P_f(\theta_m), P_{\phi}(\theta_m)$ are classical parameters.
The $U_e$ gate puts qubit $\ket{e}$ in the $\ket{0}$ state with probability $\Delta^{(m)}$ and the $\ket{1}$ state with probability $1-\Delta^{(m)}$.
With $\ket{e}$ initially in the $\ket{0}$ state, this entails applying a rotation
\begin{align}
    U_e^{(m)} \,=\, \left(\begin{matrix}\sqrt{\Delta^{(m)}} & -\sqrt{1-\Delta^{(m)}}\\
                                 \sqrt{1-\Delta^{(m)}}& \sqrt{\Delta^{(m)}}\end{matrix}\right) \,.
\end{align}
As $\Delta^{(m)}$ depends on the $\ket{n_a}, \ket{n_b}, \ket{n_{\phi}}$, this rotation must be controlled on all three count registers.
As $\Delta^{(m)}$ takes a different value for each combination of counts, we control on every possible combination, of which there are $\bO\left((m+n_I)^3\right)$.
Each rotation is controlled on $3\ceil{\log_2(n_I+m+1)}$ qubits.
A standard method \cite{nielsen_chuang_2010} decomposes multi-control gates into a sequence of Toffoli gates, which can each be implemented using six CNOTs.
Then the total cost to implement a single $3\ceil{\log_2(n_I+m+1)}$-controlled rotation is
\begin{align}
    12\cdot\Big(3\ceil{\log_2(n_I+m+1)}-1\Big)
\end{align}
CNOTs and
\begin{align}
    3\ceil{\log_2(n_I+m+1)}-1
\end{align}
re-usable ancillary qubits.

\subsubsection{\texorpdfstring{$U_{h}$}{Uh} history update}
\label{subsubsec:Uh_old}
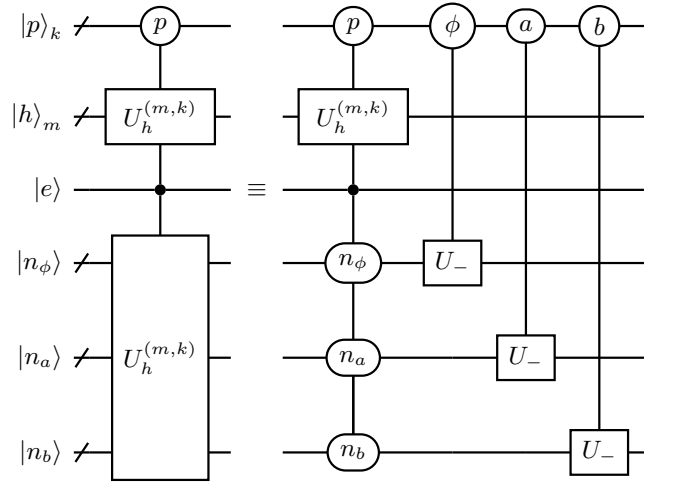
\begin{figure}[htp]
  \centering
  \begin{adjustbox}{width=1.0\columnwidth}
    \begin{tikzcd}[column sep = 0.2cm]
        \lstick{$\ket{p}_k$} & \qwbundle \qw & \measure{\mbox{$p$}} \vqw{2} & \qw & & & \measure{\mbox{$p$}} \vqw{2} & \measure{\mbox{$\phi$}} \vqw{3} & \measure{\mbox{$a$}} \vqw{4}    & \measure{\mbox{$b$}} \vqw{5} & \qw \\
        \lstick{$\ket{h}_m$} & \qwbundle \qw & \gate{U^{(m,k)}_{h}}{1} & \qw & & & \gate{U^{(m,k)}_{h}}{1} & \qw & \qw & \qw & \qw \\ 
        \lstick{\ket{e}} & \qw & \ctrl{1} & \qw & \equiv & & \ctrl{1} & \qw & \qw & \qw & \qw \\
        \lstick{\ket{n_{\phi}}} & \qwbundle \qw & \gate[3]{U^{(m,k)}_{h}} &  \qw & & & \measure{\mbox{$n_\phi$}} \vqw{1} & \gate{U_-} & \qw & \qw & \qw \\
        \lstick{\ket{n_a}} & \qwbundle \qw & \ghost{U^{(m,k)}_{h}} & \qw & & & \measure{\mbox{$n_a$}} \vqw{1} & \qw & \gate{U_-}& \qw & \qw \\ 
        \lstick{\ket{n_b}} & \qwbundle \qw & \ghost{U^{(m,k)}_{h}} & \qw & & & \measure{\mbox{$n_b$}} \vqw{-1} & \qw & \qw & \gate{U_-} & \qw   
    \end{tikzcd}
  \end{adjustbox}
  \caption{A rotation-decrement iterate used to construct $U_h$.}
  \label{fig:Uh_old_iterate}
\end{figure}

\begin{figure}[htp]
  \centering
  \begin{adjustbox}{width=1.0\columnwidth}
    \begin{tikzcd}[column sep = 0.1cm]
        \lstick{$\ket{p}_{n_I+m}$} & \qwbundle \qw & \qw & \qw & \qw & \qw & \qw & & & \measure{\mbox{$p$}} \vqw{5} & \qw & \qw \\
        &&&&&&& \dotsc &&&& \\
        \lstick{$\ket{p}_2$} & \qwbundle \qw & \qw & \qw & \qw & \measure{\mbox{$p$}} \vqw{3} & \qw & & & \qw & \qw & \qw \\
        \lstick{$\ket{p}_1$} & \qwbundle \qw & \qw & \measure{\mbox{$p$}} \vqw{2} & \qw & \qw & \qw & & & \qw & \qw & \qw \\
        \lstick{$\ket{h}_m$} & \qwbundle \qw & \qw & \gate{U^{(m,1)}_{h}} & \qw &  \gate{U^{(m,2)}_{h}} & \qw & & & \gate{U^{(m,M)}_{h}} & \measure{\mbox{$\slash\!\!\!0$}} \vqw{1} & \qw \\
        \lstick{\ket{e}} & \qw & \qw & \ctrl{1} & \qw & \ctrl{1} & \qw & \dotsc & & \ctrl{1} & \gate{X} & \qw \\
        \lstick{\ket{n_{\phi}}} & \qwbundle \qw & \qw & \gate[3]{U^{(m,1)}_{h}} & \qw & \gate[3]{U^{(m,2)}_{h}} & \qw &  \dotsc& & \gate[3]{U^{(m,n_I+m)}_{h}} & \qw & \qw \\
        \lstick{\ket{n_a}} & \qwbundle \qw & \qw & \ghost{U^{(m,1)}_{h}} & \qw &   \ghost{U^{(m,2)}_{h}} & \qw & \dotsc & & \ghost{U^{(m,M)}_{h}} & \qw & \qw \\
        \lstick{\ket{n_b}} & \qwbundle \qw & \qw & \ghost{U^{(m,1)}_{h}} & \qw &  \ghost{U^{(m,2)}_{h}} & \qw & \dotsc & & \ghost{U^{(m,n_I+m)}_{h}} & \qw & \qw
    \end{tikzcd}
  \end{adjustbox}
  \caption{Original $U_h$ gate. Each sub-gate is defined by \protect{\cref{fig:Uh_old_iterate}}}
  \label{fig:Uh_old}
\end{figure}
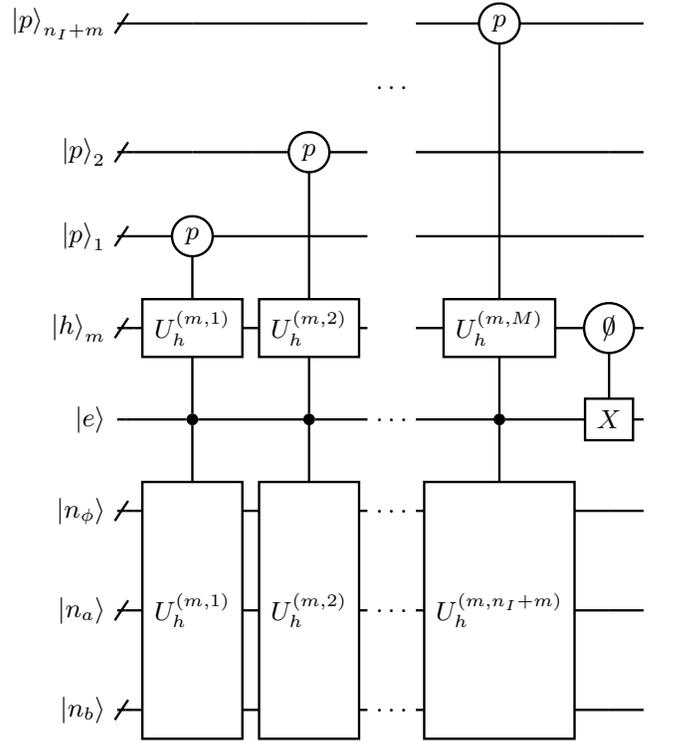

The $U_h$ gate encodes the probability of specific particles emitting. 
If there is no emission, then $\ket{h}\rightarrow\ket{0}$ with amplitude $\sqrt{\Delta^{(m)}}$
But if there is an emission, $U_h$ maps $\ket{h}$ to a superposition of basis states that correspond to particle numbers $1$ to $n_I+m$.
Each state $\ket{1}$ through $\ket{n_I+m}$ gains amplitude equal to the square root probability that the corresponding particle emits.
In particular, the probability for a specific particle to emit is equal to the total emission probability $1-\Delta^{(m)}$ times the conditional emission probability for that particle type.
Denote the relative emission probabilities
\begin{align}
    P_a(\theta_m)\,&=\,1-\Delta_a(\theta_m)\\
    P_b(\theta_m)\,&=\,1-\Delta_b(\theta_m)\\
    P_{\phi}(\theta_m)\,&=\,1-\Delta_{\phi}(\theta_m) \,.
\end{align}
Given that an emission occurred, the conditional probability that particle $p$ is the emitter is
\begin{align}
 \begin{split}
    \text{Pr}(p\,&\text{emits}\,|\,\ket{e=1}) \\
    &\,=\,\frac{P_p(\theta_m)}{\sum_{p\in\{a,b,\phi\}} n_pP_p(\theta_m)}\\
    &\,=\,\frac{P_p(\theta_m)}{\sum_{j=1}^{n_I+m}P_j(\theta_m)} \,.
    \label{eq:Uh_rel_probs}
 \end{split}
\end{align}
Therefore, $U_h$ prepares the computational basis states of $\ket{h}$ with the following amplitude distribution:
\begin{align}
    \ket{h} \mapsto &\sqrt{\Delta^{(m)}}\ket{0} + \sqrt{1-\Delta^{(m)}} \nonumber\\ &\times\sum_{j=1}^{n_I+m}\ket{j}\cdot\text{Pr}\left(p_j\,\text{emits}\,|\,\ket{e=1}\right)
    \,.
\end{align}
Starting with $\ket{h}$ in the $\ket{0}$ state, this distribution is prepared by applying a series of two-level rotations from $\ket{0}$ to the other computational basis states of $\ket{h}$.
Each rotation is controlled on $\ket{e}$, and applies the conditional emission probability for each particle $\ket{p_1}\dotsc \ket{p_{n_I+m}}$.
The rotation controlled on $\ket{p_j}$ is 
\begin{align}
    U_h^{(m,k)} \,=\, \left(\begin{matrix}\sqrt{1-\frac{P_j(\theta_m)}{\sum_{k\geq j}P_k(\theta_m)}}&-\sqrt{\frac{P_j(\theta_m)}{\sum_{k\geq j}P_k(\theta_m)}}\\
                            \sqrt{\frac{P_j(\theta_m)}{\sum_{k\geq j}P_k(\theta_m)}}&\sqrt{1-\frac{P_j(\theta_m)}{\sum_{k\geq j}P_k(\theta_m)}}\end{matrix}\right)
                            \,.
 \label{eq:Uh_rotation}
\end{align}
As the rotation angle depends on counts $\ket{n_a}, \ket{n_b}, \ket{n_{\phi}}$, rotations are also controlled on the count registers, in addition to $\ket{e}$ and $\ket{p}$.
Iterating through each particle $p_1\dotsc p_{n_I+m}$, every rotation is followed by a decrement to $\ket{n_a}$, $\ket{n_b}$, or $\ket{n_{\phi}}$ controlled on the previous particle type.
This means the relative emission probabilities given by \cref{eq:Uh_rel_probs} are updated after each rotation.
(Note that the denominator in the entries of \cref{eq:Uh_rotation} is different for each rotation.) 
\Cref{fig:Uh_old_iterate} illustrates a single rotation-decrement iterate, and \cref{fig:Uh_old} illustrates the entire $U_h$ gate.
    
The gate complexity of $U_h$ is $\bO((m+n_I)^4\log_2(n_I+m)^2)$, and $3\ceil{\log_2(n_I+m+1)}+2$ re-usable ancilla qubits are required.

\subsubsection{\texorpdfstring{$U_{p}$}{Up} particle update}
\label{subsubsec:Up_old}
\begin{figure}[htp]
  \centering
  \begin{adjustbox}{width=1.0\columnwidth}
    \begin{tikzcd}
        \lstick[wires=3]{\ket{p_{m}}} & \targ  & \qw & \qw  & \qw & \qw & \gate{U_r} & \qw & \octrl{2}  & \qw   \\
         &  \qw    & \qw & \qw & \gate{H}   & \qw & \octrl{1} &  \qw & \qw   \\ 
        & \qw  & \targ{} & \ctrl{3} & \ctrl{-1} & \ctrl{-2} & \ctrl{2} & \ctrl{1} & \qw \\
        \lstick[wires=3]{\ket{p_j}} & \qw & \ctrl{-1} & \qw & \qw & \qw  & \qw & \targ & \qw & \qw \\
        & \qw & \octrl{-1} & \qw & \qw & \qw & \targ &  \qw & \qw & \qw \\ 
        &   \ctrl{-5}  & \octrl{-1} & \targ & \qw  & \qw & \qw &  \qw & \qw & \qw
    \end{tikzcd}
  \end{adjustbox}
 \caption{$U_p$ consists of applying this gate, controlled on $\ket{h_m=j}$, for each $j$ from $1$ to $n_I+m$.}
 \label{fig:Up_old}
\end{figure}
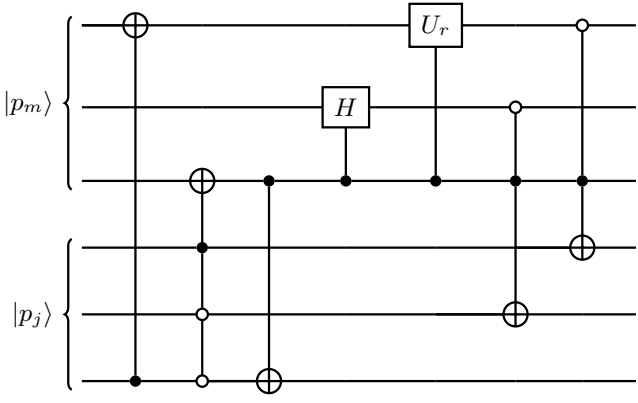

The $U_p$ gate updates the particle state $\ket{p}$ based on which particle emitted at a given step.
At step $m$, $\ket{p_m}$ stores the newly radiated particle if any.
For example, if particle $i$ emits a $\phi$ at step $m$, then $U_p$ takes
\begin{align}
    U_p\ket{p_i=f_j}\ket{p_m=0} \rightarrow \ket{f_j}\ket{\phi}
    \,.
\end{align}
Keeping in mind that $\ket{h}$ and $\ket{p}$ encode superpositions of different emission histories, $U_p$ is implemented by applying the gate in \cref{fig:Up_old} controlled on $\ket{h_m-j}$ for each $j$ from $1$ to $n_I+m$.
The single-qubit rotation $U_r$ in \cref{fig:Up_old} is the gate
\begin{align}
    U_r \,=\, \frac{1}{\sqrt{g_a^2 + g_b^2}}\left(\begin{matrix}g_a & -g_b \\ g_b & g_a\end{matrix}\right) \,.
    \label{eq:Ur}
\end{align}
In total the particle update step consists of $(n_I+m)$ applications of \cref{fig:Up_old}, each controlled on the $\ceil{\log_2(n_I+m+1)}$ qubits of $\ket{h_m}$.
The gate in \cref{fig:Up_old} has a constant number of operations, so the overall gate complexity of the particle update $U_p$ is $\bO((n_I+m)\log_2(n_I+m+1))$.

\section{Details of the QPS algorithm with mid-circuit measurements}

\subsection{\texorpdfstring{$U_{\text{count}}$}{Ucount} gate}
\label{subsubsec:Ucount_new}
We need only count the number of a single fermion type, so both gate and qubit counts are reduced by a factor of 3, compared to \cite{qps}.

The total number of CNOTs required to implement $U_{\text{count}}$ at step $m$ is 
\begin{align}
 \begin{split}
             & 13(n_{f,m})\ceil{\log_2(n_{f,m}+1)}\\
    \,\leq\, & 13(n_I+m)\ceil{\log_2(n_I+m+1)}
    \,.
 \end{split}
\end{align}

\subsection{\texorpdfstring{$U_{e}$}{Ue} gate}
\label{subsubsec:Ue_new}
Instead of conditioning on $\ket{n_a}$, $\ket{n_b}$, and $\ket{n_{\phi}}$, we only condition on $\ket{n_a}$. 
We compute Sudakov factors \cite{qps} for each possible value of $\ket{n_a}$, and apply the appropriate rotation matrices \cite{qps} conditioned on the value stored in $\ket{n_a}$.
At the $m$th simulation step, $0 \leq n_a \leq n_I + m$, so there are at most $n_I + m$ rotations, each conditioned on $\ceil{\log_2(n_I + m + 1)}$ qubits.
In the original algorithm, there are $\bO(m^3)$ rotations, which is the number of combinations of $\ket{n_a}, \ket{n_b}, \ket{n_{\phi}}$.
Here $n_{\phi}$ and $n_b$ are classically-conditioned, so the number of rotations is reduced by a factor of $m^2$.
Thus, the computational complexity of the $U_{e}$ gate is reduced to $\bO((n_I+m)\log_2\ceil{n_I + m + 1})$, compared to $\bO((n_I+m)^3\log_2\ceil{n_I + m + 1})$ for the original \cite{qps}.

\subsection{\texorpdfstring{$U_{h}$}{Uh} gate}
\label{subsubsec:Uh_new}
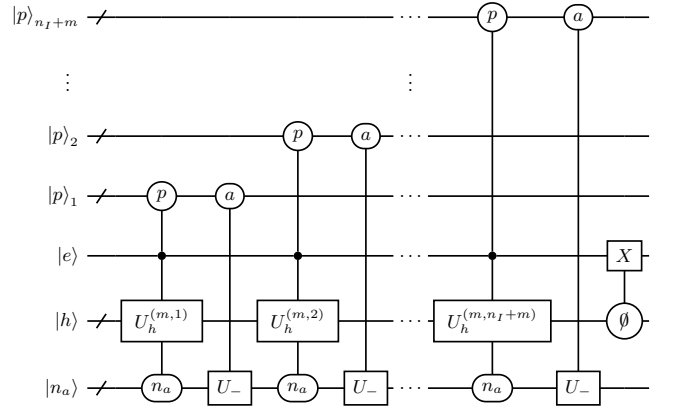
\begin{figure}[htp]
 \begin{adjustbox}{width=1.0\columnwidth}
    \begin{tikzcd}[column sep = 0.1cm]
        \lstick{$\ket{p}_{n_I+m}$} & \qwbundle \qw & \qw & \qw & \qw  &\qw  & \qw & \dotsc & & \measure{\mbox{$p$}} \vqw{4} & \measure{\mbox{$a$}} \vqw{6} & \qw & \qw \\
        \vdots\qquad&&&&&&&\vdots&&&& \\
        \lstick{$\ket{p}_2$} & \qwbundle \qw  & \qw &  \qw   & \measure{\mbox{$p$}} \vqw{2} &\measure{\mbox{$a$}} \vqw{4}  & \qw & \dotsc & & \qw & \qw & \qw & \qw      \\
        \lstick{$\ket{p}_1$} & \qwbundle \qw  &\measure{\mbox{$p$}} \vqw{1} &\measure{\mbox{$a$}} \vqw{3} & \qw   &\qw & \qw & \dotsc & & \qw & \qw & \qw & \qw \\ 
        \lstick{$\ket{e}$} & \qw &  \ctrl{1}  & \qw & \ctrl{1}  & \qw & \qw & \dots & & \ctrl{1} & \qw & \gate{X} \vqw{1} & \qw   \\
        \lstick{$\ket{h}$} & \qwbundle \qw &  \gate{U^{(m,1)}_{h}}{1} & \qw & \gate{U^{(m,2)}_{h}}{1} & \qw & \qw & \dotsc &  & \gate{U^{(m,n_I+m)}_{h}}{1} & \qw & \measure{\mbox{$\slash\!\!\!0$}} & \qw  \\
        \lstick{$\ket{n_{a}}$} & \qwbundle \qw & \measure{\mbox{$n_a$}} \vqw{-1} & \gate{U_-} & \measure{\mbox{$n_a$}} \vqw{-1} & \gate{U_-} & \qw & \dotsc & & \measure{\mbox{$n_a$}} \vqw{-1} & \gate{U_-} & \qw & \qw
    \end{tikzcd}
 \end{adjustbox}
 \caption{The $U_h$ gate at step $m$. Each $U_h^{(m,j)}$ gate denotes a sequence of two-level rotations (see \protect{\cref{subsubsec:Uh_old}}) controlled on the different states stored in $\ket{n_a}$ and $\ket{p}_j$. These controls are denoted by circular ``gates'' in the diagram. The relative probabilities for an emission from $\ket{p}_j$ depend on whether $\ket{p}_j$ is an $f_a$, $f_b$, or $\phi$, as well as the number of each particle type in the system (which is fully encoded by $\ket{n_a}$ and previous measurements of $\ket{h}$). Therefore, we must apply $3\ceil{\log_2(n_I+m+1)}$ different two-level rotations between $\ket{h=0}$ and $\ket{j}$, each conditional on the values stored in $\ket{n_a}$ and $\ket{p}_j$. Applying a controlled-decrement after each $U_h^{(m,j)}$ ensures that the correct relative emission probabilities are encoded into $\ket{h}$ (see Appendix \textcolor{red}{A.5} in \protect{\cite{qps}}), and also resets $\ket{n_a}$ to the $\ket{0}$ state. Finally, the last gate puts $\ket{e}$ back to the $\ket{0}$ state -- after updating the history register, $\ket{h}\neq\ket{0} \iff \ket{e}=\ket{1}$, so we apply a NOT gate conditional on $\ket{h}\neq\ket{0}$.}
 \label{fig:Uh_new}
\end{figure}

Like the $U_{e}$ gate, the rotations in this gate must be  conditioned only the value stored in $\ket{n_a}$.
Thus, the computational complexity of the $U_{h}$ gate is reduced to $\bO((n_I+m)^2\log_2^2\ceil{n_I + m + 1})$, compared to $\bO((n_I+m)^4\log_2^2\ceil{n_I + m + 1})$ for the original \cite{qps}.
The improved $U_h$ gate is illustrated in \cref{fig:Uh_new}. The individual $U_h^{(m,k)}$ rotations are given by \cref{eq:Uh_rotation}.

\subsection{\texorpdfstring{$U_{p}$}{Up} gate}
\label{subsubsec:Up_new}
Measuring the history register at step $m$ collapses the wavefunction of the system such that a definite emission event (including no emission) occurs with probability one.
There is still quantum interference between the different fermion types, as measuring $\ket{h}$ does not affect superpositions within particle states $\ket{p}$.

\begin{figure}[htp]
 \begin{adjustbox}{width=0.8\columnwidth}
    \begin{tikzcd}
    \lstick[wires=2]{\ket{p_{m}}} & \targ{}  & \gate{U_r} & \qw & \octrl{2} & \qw   \\
     &  \targ{} & \gate{H} & \octrl{2} &  \qw & \qw   \\
    \lstick[wires=2]{$\ket{p_{j}}$} & \targ{} & \qw & \qw   & \targ{} &  \qw & \\
     & \qw & \qw & \targ{} & \qw & \qw
    \end{tikzcd}
 \end{adjustbox}
 \caption{$U_p$ gate to be applied if $p_j=\phi$. Note that prior to the emission, $p_j$ is not encoded on qubits, so the particle update can be though of as ``initializing'' new registers $\ket{p_j}$ and $\ket{p_m}$.}
 \label{fig:Up_new}
\end{figure}
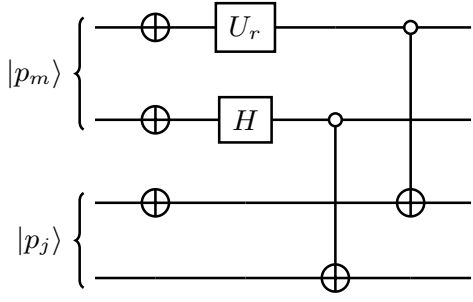

The selection of which $U_p$ gate to apply is determined dynamically based on the measurement result.
If the emitting particle is a $\phi$, then the particle update shown in \cref{fig:Up_new} is applied.

If the emitting particle is a fermion, then the particle update consists of entirely classical operations, as $\phi$ and None are not encoded on qubits.
In this case, a CPU records that particle $m$ is a $\phi$ while the emitting particle remains a fermion.
A ``particle history table'' like the one shown in \cref{tab:particle_history} is used to store the emission history as described.
The computational cost of the $U_{p}$ gate is reduced to a constant $2$ CNOTs, compared to $(n_I+m)\log_2\ceil{n_I + m + 1}$ for the original \cite{qps}.
\begin{table}[htp]
    \centering
    \begin{tabular}{c|c|c}
         Step $(m)$ & $h_m$ & Particle $m+1$   \\
         &&\\
         $0$        & $h_0$  & $\implies f/\phi$/None\\
         $1$        & $h_1$  & $\implies f/\phi$/None\\
         $2$        & $h_2$  & $\implies f/\phi$/None\\
         $3$        & $h_3$  & $\implies f/\phi$/None\\
         $\vdots$   & $\vdots$ &  $\vdots$
    \end{tabular}
    \caption{Information stored in a CPU. At each step, the history measurement determines whether particle $m+1$ will be a fermion, boson, or None. This particle history table is filled out row by row during the shower evolution, and quantum gates can be dynamically selected based on previously stored values.}
    \label{tab:particle_history}
\end{table}

\end{document}